\documentclass[twocolumn,english,aps,superscriptaddress,prb,floatfix,longbibliography]{revtex4-2}

\usepackage{color}
\usepackage{bbold}

\usepackage{float}

\usepackage{amsmath,amssymb,mathtools,physics,graphicx,bm}

\usepackage{comment}

\usepackage[unicode=true,pdfusetitle,
bookmarks=true,bookmarksnumbered=false,bookmarksopen=false,
breaklinks=true,pdfborder={0 0 0},pdfborderstyle={},backref=false,colorlinks=true]{hyperref}
\hypersetup{
colorlinks=true,
linkcolor=blue,
filecolor=magenta,      
urlcolor=blue,
citecolor=blue
}

\definecolor{cpurple}{rgb}{0.457, 0.079, 0.481}
\newcommand{\corB}[1]{\textcolor{blue}{#1}}
\newcommand{\corR}[1]{\textcolor{red}{#1}}
\newcommand{\corP}[1]{\textcolor{cpurple}{#1}}

\begin{document}
   
    \title{All-In-All-Out Pyrochlore Iridates as Noncollinear Spin-Orbit Coupled Counterparts of Altermagnets}
    \author{Yang Yang}
    \affiliation{School of Physics and Astronomy, University of Minnesota, Minneapolis, MN 55455, USA}
     \author{Turan Birol}
   \affiliation{Department of Chemical Engineering and Materials Science,
University of Minnesota, Minneapolis, MN 55455,  USA}
    \author{Rafael M. Fernandes}
\affiliation{Department of Physics, The Grainger College of Engineering,
University of Illinois Urbana-Champaign, Urbana, IL 61801, USA}
\affiliation{Anthony J. Leggett Institute for Condensed Matter Theory, The Grainger College of Engineering,
University of Illinois Urbana-Champaign, Urbana, IL 61801, USA}
  
    \author{Natalia B. Perkins}
    \affiliation{School of Physics and Astronomy, University of Minnesota, Minneapolis, MN 55455, USA}

\begin{abstract}
    Altermagnets are collinear magnetically ordered states that exhibit momentum-dependent spin splitting in the absence of net magnetization and spin-orbit coupling (SOC). Related spin-splitting patterns, however, can also emerge in noncollinear magnetic systems with large SOC. Here we show, via a microscopic model, that the all-in-all-out (AIAO)  state in pyrochlore iridates constitutes a noncollinear counterpart of a $d$-wave altermagnet stabilized by strong SOC. Starting from a microscopic $j_{\mathrm{eff}} = 1/2$ tight-binding model on the pyrochlore lattice, we demonstrate that electronic interactions favor the AIAO phase and analyze its symmetry properties. We show that the AIAO order parameter transforms as an $A_{2g}^{-}$ octupolar magnetic moment, breaking time-reversal symmetry while preserving inversion and zero net magnetization. Using group-theory analysis and mean-field calculations, we demonstrate that this symmetry enforces both a spin-polarized momentum-dependent lifting of band degeneracies that is similar to that of a collinear $d$-wave cubic altermagnet, but also a band splitting at zero-momentum. We show that the latter feature is captured by a low-energy model similar to the Luttinger-Kohn model for cubic semiconductors. Our results identify pyrochlore iridates as a platform for noncollinear counterparts of altermagnetism and provide a general symmetry framework for spin-split phenomena in spin-orbit coupled materials.
\end{abstract}
\maketitle
\section{Introduction}

Altermagnets are a class of magnetic materials that exhibit compensated collinear spin order that is invariant under the combination of time reversal and a crystal-lattice operation that includes rotations
\cite{Smejkal2022,Smejkal2022-1,Mazin2022}.
Unlike ferromagnets, which exhibit a net magnetization, or conventional antiferromagnets, which preserve spin-degenerate bands through the combined effects of time-reversal and translation or inversion, altermagnets break time-reversal symmetry without producing a net moment and give rise to spin-polarized electronic bands \cite{Jungwirth2025,Jungwirth2026symmetry}. As a result, their electronic structure exhibits characteristic nodal momentum-dependent spin splittings that cannot be described as a uniform Zeeman field.

The identification of altermagnetism as a symmetry-distinct magnetic phase followed from a classification of collinear magnetic orders in terms of spin groups, which treat spin-space and spatial symmetries independently in the absence of spin-orbit coupling (SOC) \cite{Smejkal2022,Smejkal2022-1}. This framework revealed that compensated spin configurations can generate $d$-, $g$-, or $i$-wave momentum-dependent band splittings even without net magnetization or relativistic interactions, thereby organizing collinear magnets into three symmetry-distinct classes: ferromagnets, antiferromagnets, and altermagnets. In real-space, the altermagnetic symmetries are manifested in the spin-density emerging from the magnetic atoms \cite{Jaeschke2025,Buiarelli2025}, resulting in a net ferroic order of higher-order magnetic multipoles, of which the magnetic octupole is a particular example \cite{Bhowal2022,McClarty2024,Leeb2024,Schiff2025}.

Although the spin splitting of altermagnets does not require SOC, being thus a realization of non-relativistic spin splitting \cite{Noda2016,vsmejkal2020crystal,Hayami2019,Yuan2021,Naka2020}, it is important to emphasize that SOC is compatible with altermagnetic order \cite{Fernandes2024,Scheurer2024,Antonenko2025,Agterberg2024,Roig2025,Autieri2025,Osin2026,Din2025,Campos2026}. Depending on the direction of the moments and on the underlying crystal symmetry, inclusion of SOC may result in a symmetry-allowed ferromagnetic moment (mixed altermagnets), while in other cases a net magnetization is forbidden even after SOC is included (pure altermagnets) \cite{Fernandes2024}.

There has been considerable interest in extending the concept of altermagnetism beyond collinear magnetic orders to intrinsically noncollinear spin textures \cite{Hellenes2023,Cheong2024,Liu2024,Fang2024,Song2024,Yu2025,CheongHuang2025,Liu2026symmetry}. Despite the absence of a complete symmetry framework in this case, an important question is whether noncollinear magnetic orders can exhibit the nodal momentum-dependent spin splitting characteristic of altermagnets  \cite{Cheong2024,Radaelli2024,Ghosh2026,Park2026,Hu2026}. While compensated noncollinear phases with ferroic magnetic octupolar order are natural candidates \cite{Hayamireview2024}, octupolar order alone is not sufficient. For example, the noncollinear magnetic order in Mn$_3$Sn possesses a magnetic octupole moment \cite{Nakatsuji2015,Suzuki2017}, yet does not exhibit the nodal spin splitting characteristic of altermagnets  \cite{Jungwirth2026symmetry}. Identifying noncollinear magnetic states that realize altermagnetic-like electronic structures therefore remains an open challenge.
 
\begin{figure}[t]
        \centering
        \includegraphics[width=1\linewidth]{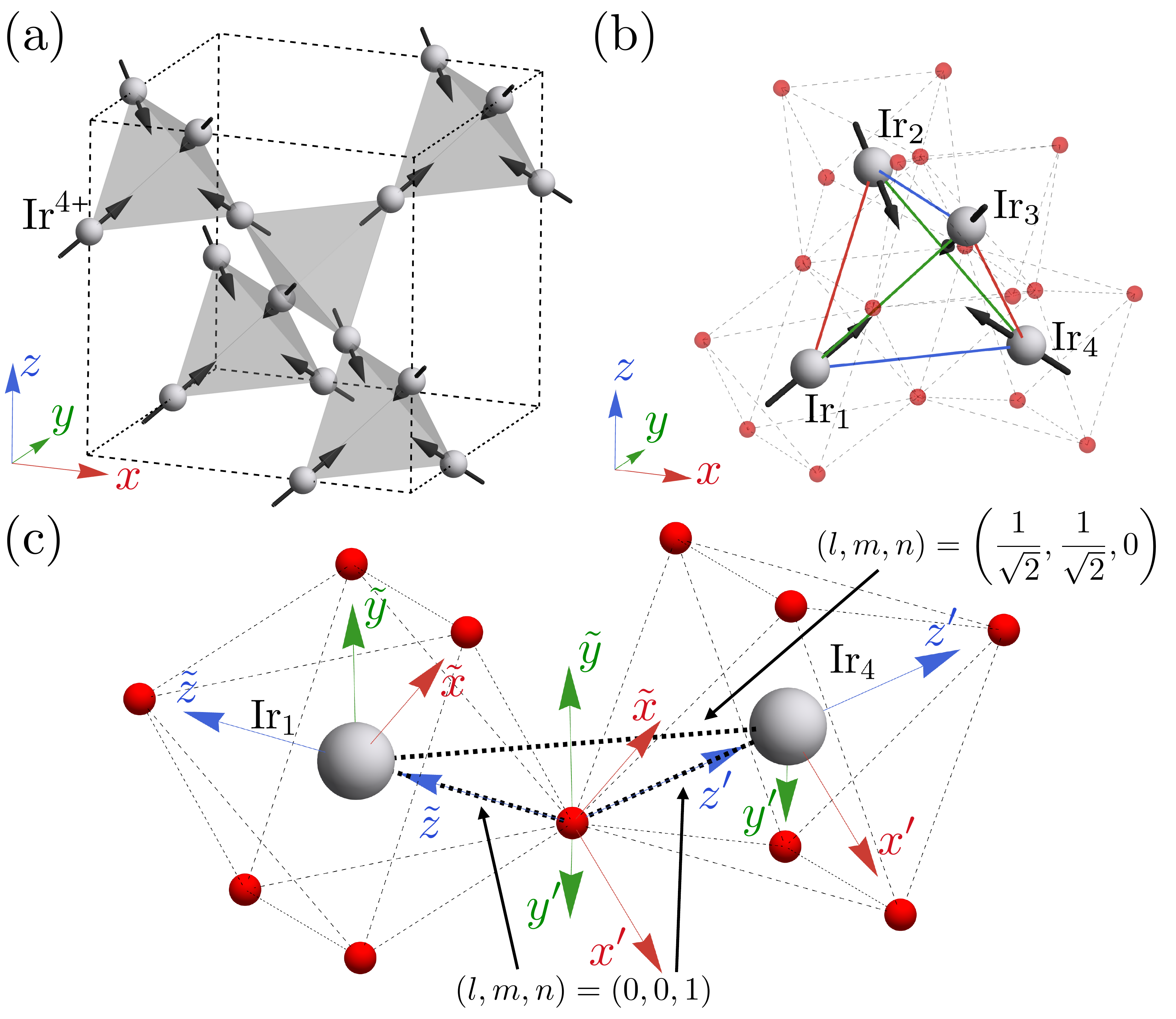}
       \caption{(a) Ir$^{4+}$ ions on the pyrochlore lattice with the all-in-all-out (AIAO) magnetic configuration. (b) The tetrahedron unit cell of the pyrochlore lattice consists of four Ir${}^{4+}$ ions surrounded by corner sharing octahedron cage of ligand ions. (c) Schematic illustration of the direct hopping and the ligand-mediated hopping between Ir$_1$ and Ir$_4$. Local coordinates $(\tilde{x},\tilde{y},\tilde{z})$ and $(x',y',z')$ are introduced to evaluate the hopping parameters. The direction cosines for Slater-Koster parameters are denoted by the unit vector $(l,m,n)$ for each pathway (see Appendix \ref{app:A}).}
        \label{fig:1}
\end{figure}

In this context, an interesting class of noncollinear magnetic materials where multipolar magnetic order is commonly observed are the  $4d$ and $5d$ transition-metal compounds with strong SOC \cite{Santini2009,Bultmark2009,voleti2020multipolar,fiore2022modeling,Urru2022}. In these cases,
spin and orbital degrees of freedom are entangled and the relevant magnetic moments are no longer pure spins but spin-orbit-entangled pseudospins \cite{Yu2024}.
 Prominent examples include the $j_{\mathrm{eff}}=1/2$ Kramers doublets that arise in Kitaev materials and related 
SOC magnets \cite{Kitaev2006,Jackeli2009,Chaloupka2010,Rau2014,Sizyuk2014,Takagi2019,Trebst2022,Rousochatzakis2024}, as well as in the pyrochlore iridates $A_2$Ir$_2$O$_7$ \cite{Pesin2010,Wan2011,WitczakKrempa2012,WitczakKrempa2013,WitczakKrempa2014,Savary2014,Goswami2017,Machida2007,Tomiyasu2012,Shapiro2012,Sagayama2013,Guo2013,Drichko2024,Xu2025}, where strong SOC and geometric frustration naturally give rise to intrinsically noncollinear magnetic order. Among these pyrochlore iridates, many members including Eu$_2$Ir$_2$O$_7$, Nd$_2$Ir$_2$O$_7$, and others, exhibit an all-in–all-out (AIAO) magnetic ground state \cite{WitczakKrempa2014,Tomiyasu2012,Shapiro2012,Sagayama2013,Guo2013,Goswami2017,Xu2025}. This state is compensated and non-collinear, as the Ir moments on each tetrahedron point either all toward or all away from its center, as shown in Fig. \ref{fig:1}(a). Moreover, it preserves translational and inversion symmetries, while displaying a net magnetic octupole moment \cite{Arima2012,Kim2020strain}. As previously pointed out \cite{Peters2024,Wu2025}, this suggests a connection between AIAO order and altermagnetism, which remains underexplored.
    
Here we combine microscopic model calculations with symmetry analysis to argue that AIAO order in pyrochlore iridates are an appealing noncollinear spin-orbit-coupled counterpart of collinear $d$-wave altermagnets in cubic lattices. First, we introduce a minimal tight-binding description of the $j_{\mathrm{eff}}=1/2$ states on the pyrochlore lattice and briefly review how electronic interactions stabilize the AIAO magnetic order, following earlier theoretical works~\cite{Pesin2010,WitczakKrempa2012,WitczakKrempa2013}. We then establish a direct relationship between the Hubbard repulsion and the AIAO order parameter,  which transforms as the $A_{2g}^-$ irreducible representation (irrep) of the $m\bar{3}m$ ($O_h$) point group. By computing the isosurfaces of the local magnetization density, we show that this order parameter is manifested, in real-space, as net magnetic octupole and dotriacontapole moments.

Our microscopic calculations reveal momentum-dependent spin splitting in the electronic band structure that is reminiscent of $d$-wave altermagnets, in that it displays nodal lines and uniaxial spin polarization along high-symmetry planes. This is to be contrasted with the nodeless spin-textures of the aforementioned noncollinear magnet Mn$_3$Sn, which also has a net magnetic octupole moment \cite{Nakatsuji2015,Suzuki2017}. 
Beyond this altermagnetic-like band splitting, however, we find an additional feature that has no counterpart in standard $S=1/2$ altermagnets. In particular, the AIAO order also generates a band splitting at the $\Gamma$ point of the Brillouin zone ($\mathbf{k}=0$). Such a $\Gamma$-point splitting does not imply a net magnetization, as it lifts a fourfold-degenerate state into two Kramers-degenerate doublets.

To understand the origin of this additional $\Gamma$-point splitting, we construct a low-energy theory based on the standard Luttinger-Kohn model for cubic semiconductors~\cite{Luttinger1955,Luttinger1956}. The fourfold-degenerate states at the $\Gamma$ point can be described in terms of an effective $J=3/2$ pseudospin. These states are not associated with atomic $J=3/2$ moments, but rather with molecular orbitals formed from coherent superpositions of $j_{\mathrm{eff}}=1/2$ states on the four Ir atoms in the unit cell.

The resulting $J=3/2$ altermagnetic low-energy theory reproduces the nodal momentum-dependent band splitting characteristic of conventional $d$-wave $S=1/2$ altermagnets. At the same time, the larger $J=3/2$ Hilbert space admits additional symmetry-allowed couplings that are absent in standard altermagnetic models, including a momentum-independent term responsible for the $\Gamma$-point band splitting revealed by our microscopic calculations. We also argue that any compensated collinear magnetic state on the pyrochlore lattice necessarily lowers the cubic symmetry by selecting a common spin axis, whereas the AIAO state preserves the cubic symmetry through its noncollinear arrangement of moments. Our work therefore identifies noncollinear spin-orbit-coupled compensated magnetic phases as the natural counterparts of collinear $d$-wave altermagnets and provides a broader framework for understanding altermagnetic-like band splitting in noncollinear magnetic systems.

The paper is organized as follows.  In Sec.~\ref{sec:model}, we introduce the interacting microscopic model that realizes AIAO order as the mean-field ground state on the pyrochlore lattice. In Sec.~\ref{sec:groundstate}, we present a symmetry analysis of the AIAO order, which we then extend in Sec.~\ref{sec:symmetry} to the resulting electronic band structure and compare with the non-magnetic
case. Finally, in Sec.~\ref{sec:A2gcouplins}, we propose a low-energy effective $J=3/2$ model to explain the origin of the band splitting, and discuss the implications of symmetry-allowed couplings. 

\section{Interacting Microscopic Model for Pyrochlore Iridates}\label{sec:model}
We consider Ir${}^{4+}$ ions ($5d{}^5$) on the pyrochlore lattice, which has cubic $O_h$ point-group symmetry (see Fig.~\ref{fig:1}). The octahedron ligand cage surrounding the Ir${}^{4+}$ ion creates a crystal field that splits the full $d$ orbital into $t_{2g}$ and $e_g$ orbitals. In the presence of spin-orbit coupling, the $t_{2g}$ orbital further splits into $j_{\mathrm{eff}}=3/2$ and $j_{\mathrm{eff}}=1/2$ orbitals. In the strong  spin-orbit coupling limit, the lower $j_{\mathrm{eff}}=3/2$ orbitals are fully occupied with four electrons, leaving the $j_{\mathrm{eff}}=1/2$ orbitals half occupied. We therefore consider a nearest-neighbor (NN) tight-binding model with onsite Hubbard repulsion $U$ involving only the $j_{\mathrm{eff}}=1/2$ orbitals (labeled by pseudospin-$\uparrow$ and -$\downarrow$) \cite{Pesin2010, WitczakKrempa2012,WitczakKrempa2014,Kurita2011} 
\begin{align}
    \mathcal{H}= U \sum_i n_{i\uparrow} n_{i\downarrow} + \sum_{ij} \sum_{mm'} \mathcal{T}_{mm'}^{ji} d_{im}^\dagger d_{jm'},\label{eqn:tightbinding}
\end{align}
where $n_{i\uparrow} = d_{i\uparrow}^\dagger d_{i\uparrow}$ and $n_{i\downarrow} = d_{i\downarrow}^\dagger d_{i\downarrow}$ are the number operators for the pseudospin-$\uparrow$ and -$\downarrow$ states, respectively, and $\mathcal{T}_{mm'}^{ji}$ denotes the hopping amplitude between sites $i$ and $j$ for pseudospin components $m$ and $m'$.
The spin dependence in the hopping matrix $\hat{\mathcal{T}}^{ji}$ originates from the different local quantization axes of the $t_{2g}$ orbitals at different sites, and becomes explicit upon projecting the hopping matrices defined in local orbital bases onto the global $j_{\mathrm{eff}}=1/2$ manifold.
 
We derive the hopping matrix $\hat{\mathcal{T}}^{ji}$  microscopically by first
considering a representative nearest-neighbor bond between Ir$_1$  and
Ir$_4$ [see Fig.~\ref{fig:1} (c)]. The hopping matrices on all other bonds follow by symmetry. The hopping between the local
      $t_{2g}$ orbitals ($d_{\tilde{y}\tilde{z}}$, $d_{\tilde{z}\tilde{x}}$, $d_{\tilde{x}\tilde{y}}$) on the $\mathrm{Ir}_1$ site 
      and ($d_{y'z'}$, $d_{z'x'}$, $d_{x'y'}$) 
      on the $\mathrm{Ir}_4$ site, where tildes and primes denote orbital components defined with respect to the distinct local coordinate frames of the two sites, consists of two 
physically distinct contributions.
The first contribution arises from direct overlap between neighboring Ir
$d$ orbitals and is parameterized by the Slater-Koster integrals
$dd\sigma$, $dd\pi$, and $dd\delta$ \cite{SlaterKoster1954}. The second contribution is
ligand-mediated and originates from second-order hopping
processes through intermediate ligand $p$ orbitals, characterized by
the hybridization amplitude $pd\pi$. Once the relative orientations of
the local orbital frames are specified, both contributions can be
evaluated systematically.
The full microscopic derivation, including explicit expressions for Slater-Koster
matrices, coordinate transformations, and projection onto the global
$j_{\mathrm{eff}}=1/2$ manifold, is presented in Appendix~\ref{app:A}.

After performing the required transformations between local and global
coordinate frames, the resulting nearest-neighbor hopping naturally
separates into an orbital and a spinor part. This separation reflects
the fact that the orbital overlap is governed by the geometry of the
local $t_{2g}$ orbitals, while the $j_{\mathrm{eff}}=1/2$ pseudospin
degrees of freedom are defined with respect to site-dependent local
quantization axes set by the IrO$_6$ octahedra. Accordingly, the hopping
between sites $i$ and $j$ may be written schematically as
\begin{align}
    \tilde{\mathcal{T}}^{ji}_{\mathrm{total}}
    =
    \tilde{\mathcal{T}}^{ji}_{\mathrm{orbital}}
    \otimes
    \tilde{\mathcal{T}}^{ji}_{\mathrm{spin}},
\end{align}
where $\tilde{\mathcal{T}}^{ji}_{\mathrm{orbital}}$ encodes the Slater-Koster
overlap of the $t_{2g}$ orbitals and
$\tilde{\mathcal{T}}^{ji}_{\mathrm{spin}}$ represents the $SU(2)$ rotation
that relates the local pseudospin frames on the two sites.
Importantly, this construction ensures that the resulting hopping
respects all lattice symmetries and incorporates spin-orbit coupling at
the microscopic level, rather than through phenomenological terms.

Projecting this structure onto the local $j_{\mathrm{eff}}=1/2$ basis
yields a hopping matrix $\tilde{\mathcal{T}}^{ji}_{J}$, in which the
orbital geometry fixes the relative amplitudes while the spin dependence
is fully encoded in the rotations between local pseudospin frames.
These rotations are described by site-dependent $SU(2)$ matrices
$\mathcal{D}_i$, which relate the local $j_{\mathrm{eff}}=1/2$
quantization axis on site $i$ to the global spin frame.
Rotating back to the global $j_{\mathrm{eff}}=1/2$ basis then gives
\begin{align}
    \mathcal{T}^{ji}=\mathcal{D}_i \tilde{\mathcal{T}}^{ji}_{J}
    \mathcal{D}_j^\dagger,
\end{align}
resulting in the compact and symmetry-constrained form
\begin{align}\label{eqn:hoppingGlobal}
    \mathcal{T}^{ji}=t\,\mathbb{I}_2 + i t'\,\mathbf{d}_{ji}\cdot
    \boldsymbol{\sigma},
\end{align}
where $\mathbb{I}_2$ is the $2\times2$ identity matrix,
$\boldsymbol{\sigma}$ are Pauli matrices, and the vectors
$\mathbf{d}_{ji}$ coincide with the Dzyaloshinskii-Moriya directions of
the pyrochlore lattice (see App.~\ref{app:A} for explicit expressions). The parameters $t$ and $t'$ encode the combined
effects of direct and ligand-mediated hopping processes, and reproduce
the standard nearest-neighbor Hamiltonian for pyrochlore iridates
obtained in
Refs.~\cite{Pesin2010,Kurita2011,WitczakKrempa2012,WitczakKrempa2013}. Physically, 
$t$ primarily controls the overall bandwidth, while 
$t'$ captures the SOC-induced anisotropic hopping responsible for momentum-dependent spin textures.

\section{Magnetic Ground State and Symmetry Properties}\label{sec:groundstate}    
        
\subsection{Mean-field ground state}

The phase diagram of the model Hamiltonian in
Eq.~\eqref{eqn:tightbinding} has been studied extensively in
Refs.~\cite{WitczakKrempa2012,WitczakKrempa2013}. For concreteness, we
choose a representative parameter set
$(U,dd\sigma,dd\pi,dd\delta,pd\pi^2/\Delta_{pd})=(0.56,-1.0,0.8,0.0,1.0)$, with all energies expressed in units of the oxygen-mediated hopping $pd\pi^2/\Delta_{pd}$. This parameter set
lies in the metallic regime of the phase diagram and yields an
AIAO magnetic ground state within self-consistent
mean-field theory. However, the conclusions below do not depend sensitively on the precise choice of parameters as long as the system remains in the AIAO phase.

The pyrochlore lattice possesses the space group $Fd\bar{3}m$ (No. 227). In the AIAO phase, the four $j_{\mathrm{eff}}=1/2$ moments on each
pyrochlore tetrahedron [Fig.~\ref{fig:1}(a)] point either all toward or all away from the
tetrahedron center, resulting in zero net magnetization per unit cell.
This noncollinear compensated therefore breaks time-reversal symmetry
while preserving the translational symmetry of the lattice.

\subsection{AIAO order as the noncollinear counterpart of a $d$-wave altermagnetic state}

Previous works have pointed out the similarity between the AIAO order and altermagnetism, as both display ferroic magnetic octupolar oder \cite{Peters2024,Wu2025}. Here, to shed further light on the consequences of the AIAO order on the electronic structure and to establish its altermagnetic-like character, we analyze its symmetry properties within the magnetic space-group framework, as appropriate for systems with sizable SOC.  We begin
by specifying the spin directions for an ``all-in'' tetrahedron [Fig.~\ref{fig:1} (b)] with
vertices at
\begin{align}
\mathbf{r}_1 &= (0,0,0), &
\mathbf{r}_2 &= \left(0,\tfrac12,\tfrac12\right), \nonumber\\
\mathbf{r}_3 &= \left(\tfrac12,0,\tfrac12\right), &
\mathbf{r}_4 &= \left(\tfrac12,\tfrac12,0\right),
\end{align}
for which the local moment orientations are
\begin{align}
\hat{\mathbf{e}}^1 &= \tfrac{1}{\sqrt{3}}(1,1,1), &
\hat{\mathbf{e}}^2 &= \tfrac{1}{\sqrt{3}}(1,-1,-1), \nonumber\\
\hat{\mathbf{e}}^3 &= \tfrac{1}{\sqrt{3}}(-1,1,-1), &
\hat{\mathbf{e}}^4 &= \tfrac{1}{\sqrt{3}}(-1,-1,1).
\label{eqn:AIAO}
\end{align}

\begin{center}
        \begin{table}[H]
            \centering
            \begin{tabular}{|c|c|c|}
                \hline
                $m\bar{3}m$   & $Fd\bar{3}m$                                                 & $\chi(A_{2g})$\\\hline
                $E$           & $\{1|(0,0,0)\}$                                              &$+1$\\\hline
                $8C_3$        & $\{3^+_{111}|(0,0,0)\}$                                      &$+1$\\\hline
                $3C_2$        & $\{2^+_{001}|\left(\frac{1}{2},\frac{1}{2},0\right)\}$       &$+1$\\\hline
                $6C_4$        & $\{4^+_{001}|\left(0,\frac{1}{2},\frac{1}{2}\right)\}$       &$-1$\\\hline
                $6C_2'$       & $\{2^+_{1\bar{1}0}|(0,0,0)\}$                                &$-1$\\\hline
                $\mathcal{I}$ & $\{\bar{1}|(0,0,0)\}$                                        &$+1$\\\hline
                $3\sigma_h$   & $\{m_{001}|\left(\frac{1}{2},\frac{1}{2},0\right)\}$         &$+1$\\\hline
                $8S_6$        & $\{\bar{3}^+_{111}|(0,0,0)\}$                                &$+1$\\\hline
                $6S_4$        & $\{\bar{4}^+_{001}|\left(0,\frac{1}{2},\frac{1}{2}\right)\}$ &$-1$\\\hline
                $6\sigma_d$   & $\{m_{1\bar{1}0}|(0,0,0)\}$                                  &$-1$\\\hline       
            \end{tabular}
            \caption{\label{tbl:characters} Characters of the AIAO state under representative symmetry operations $\{R|\mathbf{v}\}$ of space group $Fd\bar{3}m$, grouped by conjugacy classes of the point group $m\bar{3}m$, belonging to the $A_{2g}$ irreducible representation of $m\bar{3}m$.}
        \end{table}
\end{center}

Space group operations acting on the AIAO state, defined on a single tetrahedron, fall into equivalence classes in one-to-one correspondence with the conjugacy classes of the point group $m\bar{3}m$. In particular, operations combining inversion with a translation act on the spins identically to pure inversion, so the inversion $\mathcal{I}$ about Ir${}_1$  serves as a representative of this class. Applying the representative operations (in Seitz notation $\{R|\mathbf{v}\}$, where $R$ denotes a point-group operation and $\mathbf{v}$, a translation within the unit cell) from each equivalence class to the AIAO state yields the characters: operations leaving the state invariant give $+1$, while those flipping the spins give $-1$. The results are summarized in Table~\ref{tbl:characters}, from which we identify that the AIAO state transforms under the 
 $A_{2g}^-$ irrep, thus lowering the 
space-group symmetry. 
 The $A_{2g}^-$ character of the magnetic order parameter indicates that the AIAO state preserves inversion symmetry while breaking time-reversal symmetry $\mathcal{T}$, since the superscript ``$-$''  denotes that the irrep is odd under $\mathcal{T}$. 

Despite breaking $\mathcal{T}$, the AIAO configuration remains invariant under combined operations of the form $\mathcal{T}\{R|\mathbf{v}\}$, provided that the character of $\{R|\mathbf{v}\}$ is $-1$ in Table  \ref{tbl:characters}. The symmetry operations that satisfy this criterion are a proper or improper fourfold rotation around the main axes followed by a half-translation, a twofold in-plane rotation, or a reflection with respect to a diagonal mirror.  Accordingly, the magnetic space group of the AIAO phase contains anti-unitary symmetry elements of the form
$\mathcal{T}\{R|\mathbf{v}\}$. Taken together, these symmetry properties imply that the AIAO phase is
described by the $Fd\bar{3}m'$ (No. 227.131) magnetic space group derived from the
paramagnetic $Fd\bar{3}m1'$ (No. 227.129) space group. Moreover, they also imply the absence of a net magnetization and the presence of a net magnetic octupole moment, since fourfold rotations around any of the main axes (followed by a half-translation along the transverse plane) can be combined with time reversal to leave the system invariant.

The fact that the magnetic sublattices are related by a crystalline symmetry operation that contains fourfold rotations is characteristic of $d$-wave altermagnets \cite{Antonenko2025,Agterberg2024}. In the presence of SOC, as we assumed in our analysis here, the crystalline operation acts on both the direction of the moments and the atomic positions. Therefore, the noncollinear arrangement is essential to ensure that the system is invariant under a combination of rotations and time-reversal. If we enforced a collinear magnetic order on the four-Ir sites, the moments' common axis would necessarily break the cubic symmetry, which must be preserved for a magnetic state described by an $A_{2g}^-$ (i.e., magnetic multipole) order parameter.  The fact that a $d$-wave collinear altermagnetic order cannot be realized in the pyrochlore lattice is consistent with the results of Ref. \cite{Schiff2025}, which show that magnetic atoms on the Wyckoff position $16d$ of space group $Fd\bar{3}m$ (No. 227), which is the position of the Ir atoms, do not support an altermagnetic state. 

In the following sections, we
analyze how this $A^-_{2g}$ altermagnetic-like order parameter couples to the electronic
degrees of freedom and lifts symmetry-protected band degeneracies, in analogy with collinear altermagnets.\\

\section{Symmetry Analysis of the Electronic Band structure}\label{sec:symmetry}   

\subsection{Non-magnetic band structure}\label{nonmagnetic}

\begin{figure}[t]
            \centering
            \includegraphics[width=1\linewidth]{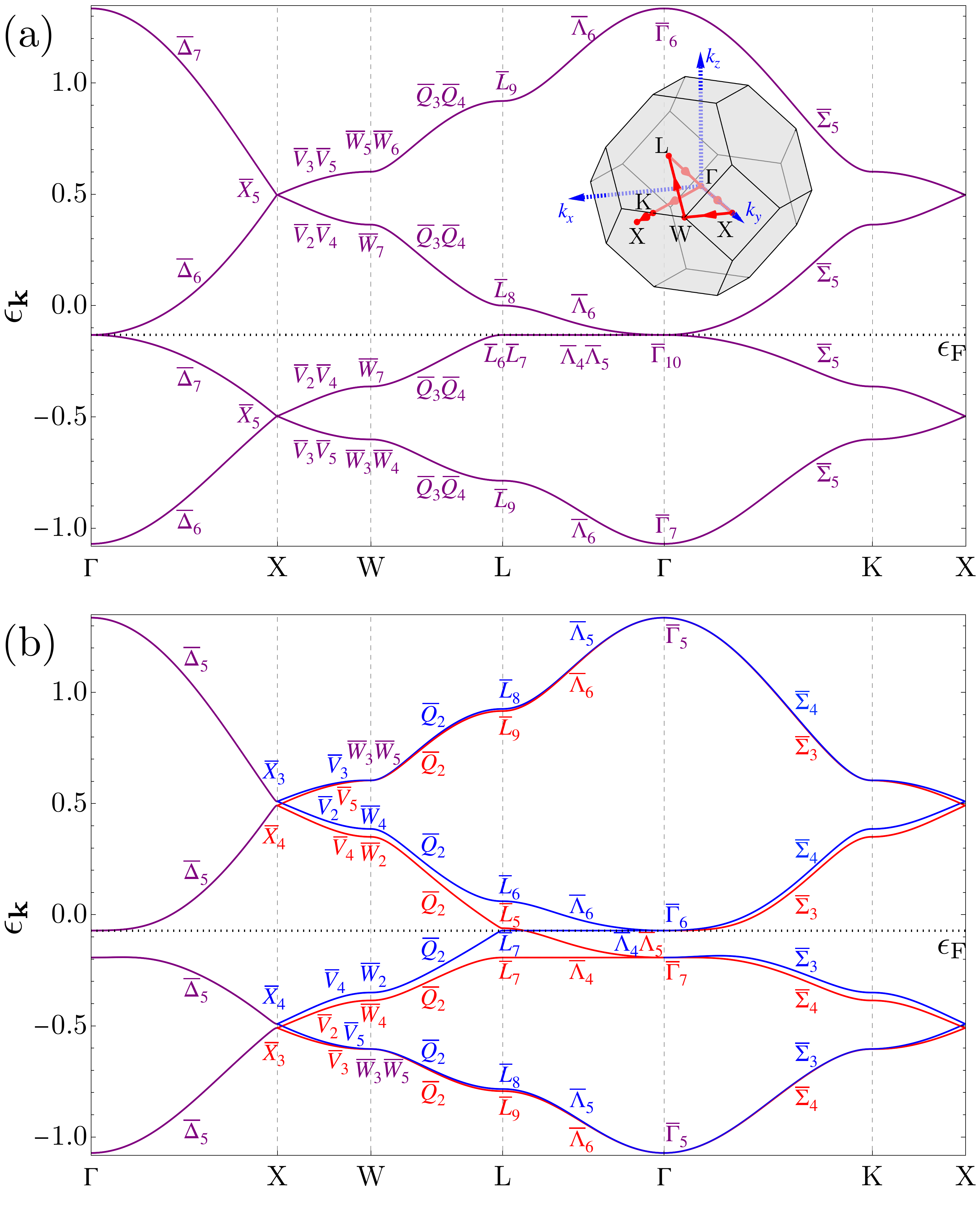}
            \caption{Pseudospin-splitting of the band structure of pyrochlore iridates in the presence of AIAO order. (a) Band structure along high-symmetry lines (inset shows the path taken in momentum space) without magnetic ordering computed with $U=0$. The bands are labeled by irreducible co-representations (co-irreps) of the magnetic space group $Fd\bar{3}m1'$ (No. 227.129), see Table \ref{tbl:coirrep}. The spectrum is at least two-fold degenerate everywhere, and thus colored purple. (b) Corresponding band structure with AIAO order, obtained from the mean-field self-consistent calculation with $U=0.56$ in units of $pd\pi^2/\Delta_{pd}$. The two pseudospin-split bands are colored blue and red and labeled according to the induced co-irreps ${}^1\overline{E}_g\uparrow Fd\bar{3}m'$ (red) and ${}^2\overline{E}_g\uparrow Fd\bar{3}m'$ (blue) at Wyckoff position $16d$ for magnetic space group $Fd\bar{3}m'$ (No. 227.131). The non-split bands remain purple. }
            \label{fig:spectrum}
        \end{figure}

We begin by establishing the symmetry and degeneracy of the electronic bands in the absence of magnetic order, which serves as a reference  for understanding the spin splitting induced by AIAO ordering. We therefore set 
$U=0$
and compute the non-magnetic band structure of the tight-binding Hamiltonian in Eq.~\eqref{eqn:tightbinding} with $(dd\sigma,dd\pi,dd\delta,pd\pi^2/\Delta_{pd})=(-1.0,0.8,0.0,1.0)$. The results are shown in Fig.~\ref{fig:spectrum} (a).

In this regime, the system preserves time-reversal symmetry as well as all spatial symmetries of the pyrochlore lattice, including inversion. As a result, the electronic bands are at least twofold degenerate throughout the Brillouin zone due to Kramers degeneracy, becoming fourfold degenerate at certain high-symmetry points.
Since the non-magnetic ground state is described by the (para)magnetic space group $Fd\bar{3}m1'$, all eigenstates of
 Eq.~\eqref{eqn:tightbinding}
must transform according to the irreducible co-representations (co-irreps) of this group, which we use to label the electronic bands. These co-irreps act as extensions of the parent space group's irreducible representations, explicitly accounting for the antiunitary operations.

The co-irreps carried by the bands can be deduced from the site symmetry of the Wyckoff position that hosts the relevant electronic degrees of freedom \cite{Evarestov2012}. In our system, these degrees of freedom reside on the $\mathrm{Ir}^{4+}$ ions at the $16d$ Wyckoff position of $Fd\bar{3}m1'$. The corresponding site symmetry is the magnetic point group $\bar{3}m1'$. Under this group, the local electronic degrees of freedom described by $j_{\mathrm{eff}}=1/2$ states have to transform as the co-irrep obtained from group subduction $D^{1/2} \downarrow \bar{3}m1' = \overline{E}_{1g}$ \cite{Inui2012}. This can be explicitly verified by checking the characters of the symmetry operations from $\bar{3}m1'$ applied to $j_{\mathrm{eff}}=1/2$ states. The co-irreps for the electronic bands across the Brillouin zone are then obtained from the induced co-irrep $\overline{E}_{1g} \uparrow Fd\bar{3}m1'$ evaluated at various $\mathbf{k}$-points. The resulting band structure along high-symmetry lines and points, with bands labeled according to the co-irreps listed in the first column of Table~\ref{tbl:coirrep}, are shown in Fig.~\ref{fig:spectrum}(a). Note that some of the induced co-irreps at high symmetry points, such as $\overline X_5$ and $\overline\Gamma_{10}$, have a dimension $4$, corresponding to the fourfold degenerate states mentioned above.

 \begin{widetext}
    \begin{center}
    \begin{table}[H]
        \centering
        \begin{tabular}{|c|c|c|c|c|}
            \hline
                    &$(k_x,k_y,k_z)$& $\corP{\overline{E}_{1g}}\uparrow Fd\bar{3}m1'$& $\corR{{}^1\overline{E}_g}\uparrow Fd\bar{3}m'$ &  $\corB{{}^2\overline{E}_g}\uparrow Fd\bar{3}m'$\\\hline
                    
            $\Delta$ & $(0,v,0)$& $2\overline{\Delta}_6(2)\oplus 2\overline{\Delta}_7(2)$ & $2\overline{\Delta}_5(2)$& $2\overline{\Delta}_5(2)$\\\hline
            
            $X$ & $(0,1,0)$& $2\overline{X}_5(4)$ & $\overline{X}_3(2)\oplus\overline{X}_4(2)$& $\overline{X}_3(2)\oplus\overline{X}_4(2)$\\\hline
            
            $V$ & $(u,1,0)$ &$2\overline{V}_2\overline{V}_4(2)\oplus2\overline{V}_3\overline{V}_5(2)$ & $\overline{V}_2(1)\oplus\overline{V}_3(1)\oplus\overline{V}_4(1)\oplus\overline{V}_5(1)$&$\overline{V}_2(1)\oplus\overline{V}_3(1)\oplus\overline{V}_4(1)\oplus\overline{V}_5(1)$\\\hline
            
            $W$ & $(1/2,1,0)$ & $\overline{W}_3\overline{W}_4(2)\oplus\overline{W}_5\overline{W}_6(2)\oplus 2\overline{W}_7(2)$ & $\overline{W}_2(1)\oplus\overline{W}_3\overline{W}_5(2)\oplus\overline{W}_4(1)$ & $\overline{W}_2(1)\oplus \overline{W}_3\overline{W}_5(2)\oplus\overline{W}_4(1)$\\\hline
            
            $Q$  & $(1/2,v,1-v)$ & $4\overline{Q}_3\overline{Q}_4(2)$ & $4\overline{Q}_2(1)$ & $4\overline{Q}_2(1)$\\\hline
            
            $L$ & $(1/2,1/2,1/2)$ & $\overline{L}_6\overline{L}_7(2)\oplus\overline{L}_8(2)\oplus 2\overline{L}_9(2)$ & $\overline{L}_5(1)\oplus\overline{L}_7(1)\oplus\overline{L}_8(1)\oplus\overline{L}_9(1)$ & $\overline{L}_6(1)\oplus\overline{L}_7(1)\oplus\overline{L}_8(1)\oplus\overline{L}_9(1)$\\\hline
            
            $\Lambda$ & $(u,u,u)$ &$\overline{\Lambda}_4\overline{\Lambda}_5(2)\oplus 3\overline{\Lambda}_6(2)$ & $\overline{\Lambda}_4(1)\oplus 2\overline{\Lambda}_5(1)\oplus \overline{\Lambda}_6(1)$ & $\overline{\Lambda}_4(1)\oplus\overline{\Lambda}_5(1)\oplus 2\overline{\Lambda}_6(1)$\\\hline
            
            $\Gamma$& $(0,0,0)$ &$\overline{\Gamma}_6(2)\oplus \overline{\Gamma}_7(2)\oplus \overline{\Gamma}_{10}(4)$& $\overline{\Gamma}_5(2)\oplus \overline{\Gamma}_7$(2)& $\overline{\Gamma}_5(2)\oplus \overline{\Gamma}_6$(2)\\\hline
            
            $\Sigma$ & $(u,u,0)$ &$4\overline{\Sigma}_5(2)$ &$2\overline{\Sigma}_3(1)\oplus 2\overline{\Sigma}_4(1)$ & $2\overline{\Sigma}_3(1)\oplus 2\overline{\Sigma}_4(1)$\\\hline
        \end{tabular}
        \caption{Induced irreducible co-representations (co-irreps) at Wyckoff position $16d$ for $j_{\mathrm{eff}}=1/2$ degrees of freedom in the magnetic space groups $Fd\bar{3}m1'$ (No. 227.129) and $Fd\bar{3}m'$ (No. 227.131), obtained from Bilbao Crystallographic Server \cite{Elcoro2017,Xu2020}. The site symmetry of the $\mathrm{Ir}^{4+}$ ions (Wyckoff position 16d) is $\bar{3}m1'$ in $Fd\bar{3}m1'$ and $\bar{3}m'$ in $Fd\bar{3}m'$. $\overline{E}_{1g}$ is a two-dimensional irreducible co-irrep of $\bar{3}m1'$ whereas ${}^{1}\overline{E}_{g}$ and ${}^{2}\overline{E}_{g}$ are a pair of one-dimensional irreducible co-irreps of $\bar{3}m'$. Numbers in parentheses indicate the dimension of each induced co-irrep.}
        \label{tbl:coirrep}
    \end{table}
    \end{center}
    \end{widetext}

\subsection{Band splitting in the AIAO phase}\label{AIAObandstructure}

We now turn to the electronic structure in the presence of AIAO order. To this end, we switch on the Hubbard interaction and get back to the representative parameter set 
$(U,dd\sigma,dd\pi,dd\delta,pd\pi^2/\Delta_{pd})=(0.56,-1.0,0.8,0.0,1.0)$,
known to stabilize the AIAO phase within self-consistent mean-field theory \cite{WitczakKrempa2012}, and obtain the effective spin length $S=0.1084$. The resulting band structure is shown in 
Fig.~\ref{fig:spectrum}(b), allowing for a direct comparison with the non-magnetic band structure in Fig.~\ref{fig:spectrum}(a).

As discussed above, the AIAO order lowers the system symmetry from $Fd\bar{3}m1'$ to $Fd\bar{3}m'$. Consequently, the electronic bands must therefore be classified according to the co-irreps of the latter.
This symmetry reduction has a direct and transparent impact on the low-energy $j_{\mathrm{eff}}=1/2$ states.
In the non-magnetic phase, the site-symmetry group $\bar{3}m1'$ of the $16d$ 
site requires the $j_{\mathrm{eff}}=1/2$ states to transform as the co-irrep 
$\overline{E}_{1g}$.
Because this co-irrep is two-dimensional, it enforces a 
twofold degeneracy protected by time-reversal symmetry. Upon ordering into the 
AIAO phase, the site symmetry is reduced to $\bar{3}m'$, and the compatibility 
relations dictate the decomposition 
$\overline{E}_{1g}\rightarrow{}^{1}\overline{E}_{g}\oplus{}^{2}\overline{E}_{g}$ 
into two one-dimensional co-irreps. With time-reversal symmetry now broken, the 
residual antiunitary operations no longer enforce this degeneracy, and the two 
co-irreps are therefore free to split in energy, giving rise to the pseudospin 
splitting observed in the electronic band structure.

To verify this, we obtain the band labels using induced co-irreps of the magnetic space group at the Wyckoff position $16d$. The labeling in Fig.~\ref{fig:spectrum}(a) follows from the induced co-irreps
$\overline{E}_{1g}\uparrow Fd\bar{3}m1'$ at Wyckoff position $16d$. In the AIAO phase, the bands in Fig.~\ref{fig:spectrum}(b) are instead labeled by the induced co-irreps
 ${}^1\overline{E}_g\uparrow Fd\bar{3}m'$ and ${}^2\overline{E}_g\uparrow Fd\bar{3}m'$. These induced co-irreps are obtained from the Bilbao Crystallographic Server \cite{Elcoro2017,Xu2020} and listed in Table~\ref{tbl:coirrep}.

 Viewed in this way, the pseudospin splitting of the electronic bands can be understood as a consequence of the symmetry-driven decomposition
 \begin{eqnarray}
     \overline{E}_{1g}\uparrow Fd\bar{3}m1'\rightarrow{}^1\overline{E}_g\uparrow Fd\bar{3}m'\oplus{}^2\overline{E}_g\uparrow Fd\bar{3}m',
 \end{eqnarray}
  with the two split branches shown in red and blue in Fig.~\ref{fig:spectrum}(b). Importantly, this splitting is symmetry-enforced rather than arising from a conventional Zeeman mechanism. Its origin lies in the condensation of the $A_{2g}^{-}$ AIAO order parameter, which breaks time-reversal symmetry and lowers the magnetic space-group symmetry without producing a net magnetization. As a result, the band splitting is strongly momentum dependent and non-uniform across the Brillouin zone, as dictated by the symmetry of the noncollinear AIAO magnetic state, providing a clear signature of altermagnetic-like splitting in a noncollinear compensated magnet. Indeed, from Fig.~\ref{fig:spectrum}(b), we note the absence of pseudospin-splitting along the $\Gamma$-$X$  direction, which coincides with the spin-splitting nodal line expected for a standard $d$-wave altermagnetic order parameter on the cubic lattice \cite{Fernandes2024}. This is consistent with the group-theory analysis of Table \ref{tbl:coirrep}, which shows that every state along the $\Delta$ line, which connects the  $\Gamma$ and $X$  points, is twofold degenerate with or without AIAO order. On the other hand, we also note both from Table  \ref{tbl:coirrep} and Fig. \ref{fig:spectrum} that AIAO order also causes a band splitting at the $\Gamma$-point by lowering the fourfold degeneracy of the $\overline{\Gamma}_{10}$ state to twofold degenerate split $\overline{\Gamma}_6$ and $\overline{\Gamma}_7$ states. Similarly, a band splitting associated with the fourfold degenerate $\overline{X}_5$  state is also observed. We will further discuss these splittings in Sec. \ref{sec:A2gcouplins}.

\begin{figure}[t]
    \centering
    \includegraphics[width=1\linewidth]{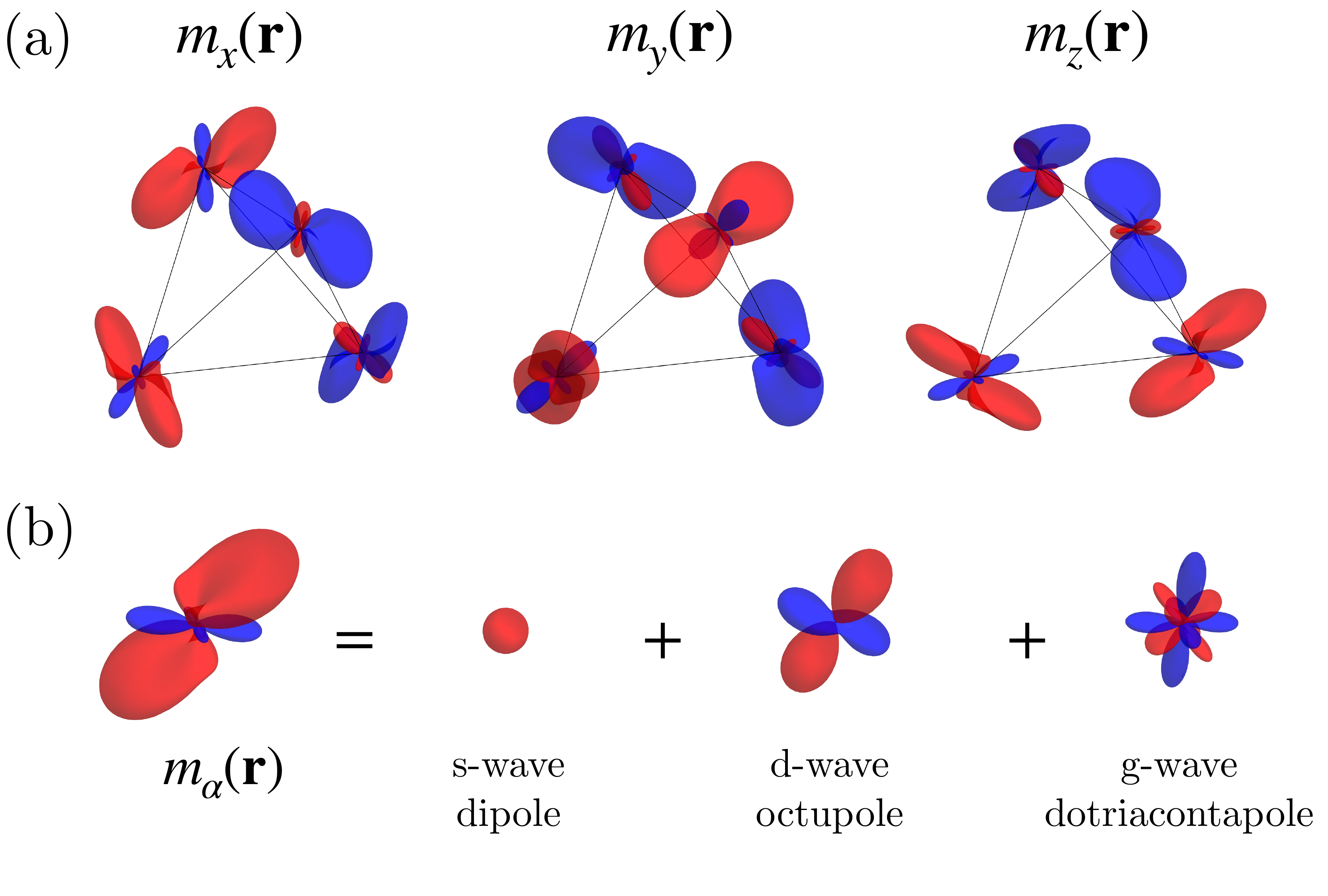}
    \caption{
    (a) Isosurfaces of local magnetization density $m_\alpha(\mathbf{r})$ ($\alpha=x,y,z$) on a tetrahedral  unit cell calculated from our microscopic model. (b) Multipole decomposition of $m_\alpha(\mathbf{r})$ into $s$-, $d$-, and $g$-wave components ($l=0,2,4$), corresponding to dipole, octupole, and dotriacontapole moments, respectively.}
    \label{fig:magnetization-density}
\end{figure}

\subsection{Real-space magnetization density and multipolar structure}\label{magnetization}

To gain further insight into the microscopic origin of the altermagnetic-like band splitting, we analyze the local magnetization density $m_\alpha(\mathbf r)$ ($\alpha = x, y, z$), defined as the expectation value of the spin–orbital magnetization density in real space (see App.~\ref{app:D} for details). In the presence of strong SOC, spin and orbital degrees of freedom are intrinsically entangled, so 
 $m_\alpha(\mathbf r)$ encodes the full spin-orbital character of the magnetic state rather than a pure spin texture.

While the band structure in Fig.~\ref{fig:spectrum} shows the typical momentum-space signature of $d$-wave altermagnetism on the cubic lattice (no pseudospin splitting along $\Gamma$-$X$), the real-space structure of $m_\alpha(\mathbf r)$ provides its microscopic origin \cite{Buiarelli2025,Jaeschke2025}. In particular, the anisotropy and multipolar structure of $m_\alpha(\mathbf r)$ directly determine the momentum-dependent pseudospin splitting observed in Fig.~\ref{fig:spectrum}(b).
As shown in Fig.~\ref{fig:magnetization-density} (a), the isosurfaces of $m_\alpha(\mathbf r)$ on the tetrahedral unit cell are strongly anisotropic, reflecting the underlying crystal symmetry. This anisotropy is further demonstrated by the multipole decomposition in Fig.~\ref{fig:magnetization-density} (b), which shows that $m_\alpha(\mathbf r)$ consists of dipole ($l=0$), octupole ($l=2$), and dotriacontapole ($l=4$) contributions, with comparable weight in the higher-order multipoles. 
 
These higher-rank multipoles encode the symmetries of the AIAO state and provide the microscopic origin of the anisotropic, momentum-dependent band splitting. For instance, while the dipole components of the magnetization densities at the four atomic positions cancel each other, the octupole components all have the same sign, demonstrating the emergent magnetic octupolar ferroic order. More broadly, the AIAO magnetic order breaks time-reversal symmetry but preserves the combination of time-reversal with other crystalline operations. This lifts the Kramers degeneracy and the resulting band splitting is a spin-orbital splitting whose momentum dependence and nodal structure are governed by the lattice point-group symmetry and the multipolar character of $m_\alpha(\mathbf r)$. 
In this sense, the multipolar structure of $m_\alpha(\mathbf r)$ provides a real-space characterization of the magnetic order, whose momentum-space manifestation is the non-uniform pseudospin splitting of the electronic bands.
In contrast to altermagnets without SOC,
where spin remains a good quantum number and the physics reduces to pure spin textures, the splitting here originates from entangled
spin-orbital multipoles, providing a natural route to understand the emergence of altermagnetic-like phenomena in noncollinear, strong-SOC magnetic materials.

\section{Low-energy effective $J=3/2$ altermagnetic model  for the AIAO phase}\label{sec:A2gcouplins}

What is the simplest effective Hamiltonian that captures the
symmetry-enforced band splitting induced by AIAO order? To address this
question, and make a more direct comparison with effective altermagnetic Hamiltonians, we construct a low-energy description of the electronic
structure for small momenta (i.e., around the $\Gamma$ point) based solely on symmetry considerations. In the absence of
magnetic order, the system is invariant under the full (para)magnetic space
group $Fd\bar{3}m1'$, and  the low-energy states near the Fermi level originate from
local $j_{\mathrm{eff}}=1/2$ Kramers doublets residing on the four Ir
sublattices of the pyrochlore unit cell. 
At the Brillouin-zone
center (the $\Gamma$ point), 
the Bloch states built from
 spin-orbit-entangled orbitals on the sublattices can be classified according to the irreducible co-representations of the magnetic point group $m\bar{3}m1'$. Since $m\bar{3}m1'$ generates the four sublattices of Ir${}^{4+}$ ions, induction to the full point group naturally incorporates the sublattice degrees of freedom. The local $j_{\mathrm{eff}}=1/2$ co-irrep
$\overline{E}_{1g}$ accordingly decomposes as
$\overline{E}_{1g}\!\uparrow\, m\bar{3}m1'
= \overline{E}_{1g}\,(\overline{\Gamma}_6)
\oplus \overline{E}_{2g}\,(\overline{\Gamma}_7)
\oplus \overline{F}_g\,(\overline{\Gamma}_{10})$,
yielding two two-dimensional representations,
$\overline{\Gamma}_6$ and $\overline{\Gamma}_7$, together with a
four-dimensional $\overline{\Gamma}_{10}$ sector
(see Table~\ref{tbl:coirrep}).

Upon the onset of AIAO order, the symmetry is reduced, and the fourfold
degeneracy of the $\overline{\Gamma}_{10}$ sector is lifted, as shown in Fig. \ref{fig:spectrum}(b) and discussed in the previous section. 
 Building on
the symmetry analysis of the preceding section, we proceed to identify the leading
symmetry-allowed couplings between the electronic states arising from  $\overline{\Gamma}_{10}$  and the $A_{2g}^{-}$ AIAO order parameter. 
The corresponding low-energy Hamiltonian must be invariant under all  symmetry
operations of the paramagnetic point group. As a result, an order parameter
transforming according to a given irrep $\Gamma$ can couple only to electronic operators that transform under the
same irrep. This observation allows the Hamiltonian describing the coupling to the AIAO order parameter to be
decomposed into symmetry-resolved channels,
\begin{align}\label{Hint}
    \mathcal{H}_{\mathrm{coupl}}=\sum_{\Gamma}\Phi^{\Gamma}
    \mathcal{H}_{\mathrm{coupl}}^{\Gamma},
\end{align}
where $\Phi^{\Gamma}$ denotes a (momentum-independent) order parameter
belonging to the irrep $\Gamma$, and
$\mathcal{H}_{\mathrm{coupl}}^{\Gamma}$ is the corresponding
electronic operator. In second-quantized form, the latter can be written in terms of fermionic bilinears as
\begin{align}\label{HintA2g}
    \mathcal{H}_{\mathrm{coupl}}^{A_{2g}^-}
    =
    \sum_{\mathbf{k}}
    \psi_{\mathbf{k}}^{\dagger}
    \, H_{A_{2g}^-}(\mathbf{k}) \,
    \psi_{\mathbf{k}},
\end{align}
where $\psi_{\mathbf{k}}^\dagger$ ($\psi_{\mathbf{k}}$) creates (annihilates) a Bloch state and $H_{A_{2g}^-}(\mathbf{k})$ is a Hermitian matrix transforming according to the $A_{2g}^{-}$ irreducible
representation. 
While such a decomposition must in principle be applied to the full microscopic Hamiltonian, we restrict it here to the low-energy $\overline{\Gamma}_{10}$ subspace, as it already captures the main fingerprints of AIAO order on the band structure.

Within the low-energy $\overline{\Gamma}_{10}$ subspace, the non-interacting electronic degrees of freedom are described by effective angular momentum $J=3/2$ operators $J_x$, $J_y$, and $J_z$,  as well established by the Luttinger-Kohn model for cubic semiconductors \cite{Luttinger1955,Luttinger1956}.  Note that these $J=3/2$ operators are unrelated to the spin-orbit coupled pseudospins $j_{\mathrm{eff}}=1/2,\,3/2$ defined in our original model. Near the $\Gamma$ point, the $\overline{\Gamma}_{6}$ and $\overline{\Gamma}_{7}$ subspaces are energetically well-separated from the $\overline{\Gamma}_{10}$ subspace, allowing us to project the full microscopic Hamiltonian \eqref{eqn:tightbinding}  onto the $\overline{\Gamma}_{10}$ subspace (see App.~\ref{app:B} for details). To quadratic order in $\mathbf{k}$, 
the non-interacting part of the microscopic Hamiltonian in Eq.~\eqref{eqn:tightbinding}, projected onto the $\overline{\Gamma}_{10}$ subspace, takes a Luttinger-Kohn form
\begin{widetext}
        \begin{align}
            \mathcal{H}^{\overline{\Gamma}_{10},(2)}_{\mathrm{hopping}}(\mathbf{k})=&\frac{t-2t'}{3}(k_x^2+k_y^2+k_z^2-6)\mathbb{1}\nonumber\\
            +&\frac{t+t'}{18}\left[3(k_x^2-k_y^2)(J_x^2-J_y^2)+(2k_z^2-k_x^2-k_y^2)(2J_z^2-J_x^2-J_y^2)\right]\nonumber\\
            -&\frac{t-2t'}{3}\left[k_yk_z(J_yJ_z+J_zJ_y)+ k_z k_x (J_xJ_z+J_zJ_x)+k_x k_y(J_xJ_y+J_yJ_x)\right]. \label{eq:H_hopping_projected}
        \end{align}
\end{widetext}
Here, $J_x$, $J_y$, and $J_z$ denote the $4\times4$ $3/2$ angular momentum matrices acting within the $\overline{\Gamma}_{10}$ subspace, and $\mathbb{1}$ is the $4\times4$ identity matrix.

\begin{figure}[t]
    \centering
    \includegraphics[width=1\linewidth]{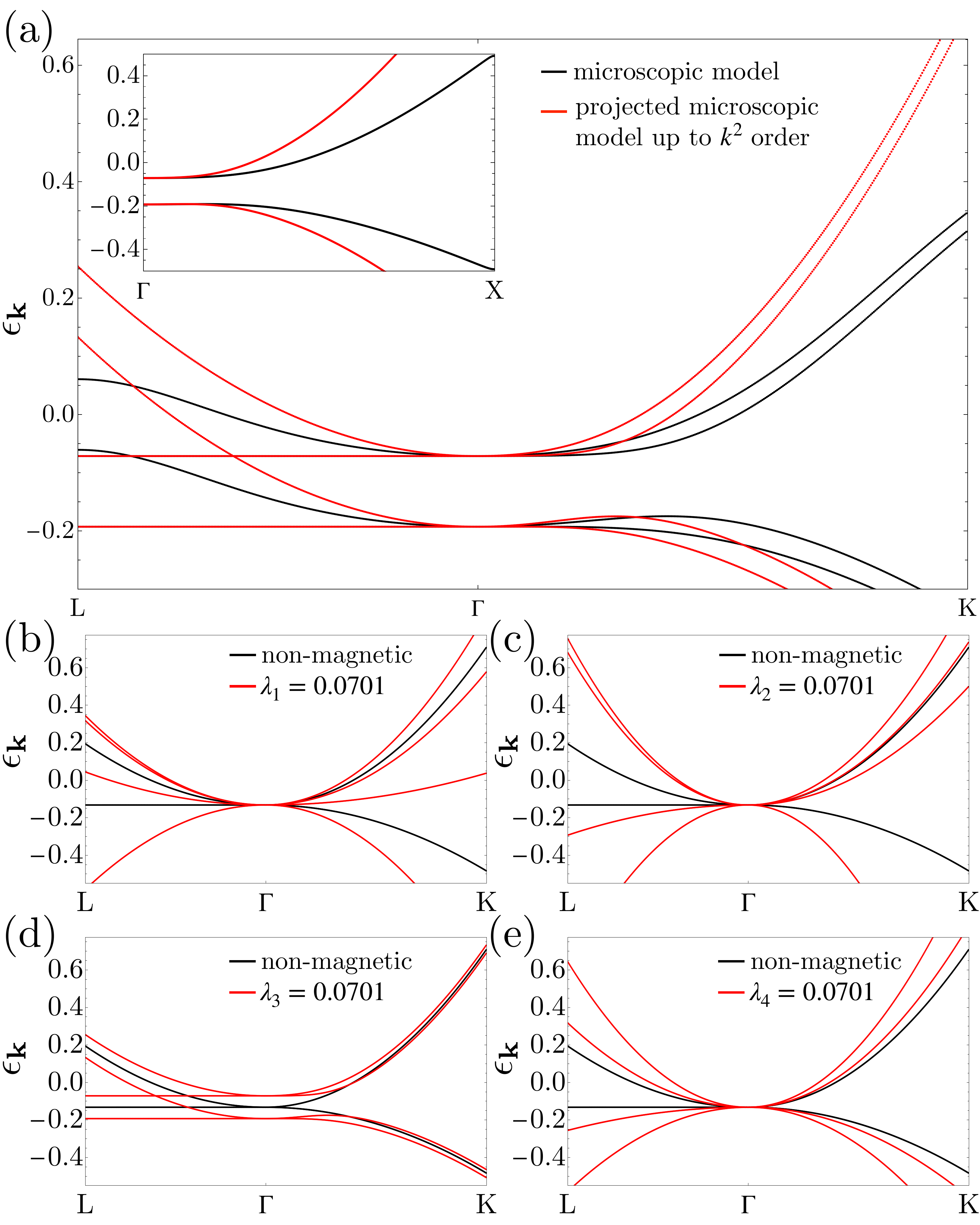}
    \caption{\label{fig:splitting}
    (a) Electronic dispersion of the $\overline{\Gamma}_{10}$ subspace along $L$-$\Gamma$-$K$ ($\Gamma$-$X$ in the inset), comparing  the full microscopic model (black) with its $\mathbf{k}^2$-order low-energy projection (red). The projected model accurately captures both the momentum-dependent band splitting induced by the AIAO order near the $\Gamma$ point and the band splitting at $\Gamma$. (b)-(e) Electronic dispersion obtained from the general low-energy Hamiltonian \eqref{eqn:symmetryA2g}, with one coupling $\lambda_\alpha$ ($\alpha=1,2,3,4$) set to a finite value while all others are set to zero, compared to the non-magnetic case (black). 
While all four terms lift the Kramers degeneracy and produce momentum-dependent splittings similar to a collinear $d$-wave altermagnetic phase, only the $\lambda_3$ term [panel (d)] splits the fourfold degenerate states at zero momentum.}
\end{figure}

Similarly, the Hubbard interaction term in \eqref{eqn:tightbinding} can be mean-field decoupled assuming AIAO ordering with effective spin directions given by Eq.~\eqref{eqn:AIAO} and spin length $S$, and then  projected onto the $\overline{\Gamma}_{10}$ subspace, yielding
\begin{align}\label{eqn:micro-interaction}
    \mathcal{H}^{\overline{\Gamma}_{10}}_{\mathrm{MF}}=\frac{2\sqrt{3}}{3}US(J_x J_y J_z+J_z J_y J_x).
\end{align}
Using the same parameters as in the self-consistent mean-field treatment of the full microscopic model, we compare the projected quadratic Hamiltonian $\mathcal{H}^{\overline{\Gamma}_{10},(2)}_{\mathrm{hopping}}(\mathbf{k})+\mathcal{H}^{\overline{\Gamma}_{10}}_{\mathrm{MF}}$ with the full microscopic model near the $\Gamma$ point,
as shown in  Fig.~\ref{fig:splitting} (a). Clearly, the projected model captures the main features of the microscopic model in the AIAO phase, namely, the momentum-dependent band splitting (including the absence of splitting along $\Gamma$-$X$ shown in the inset of Fig.~\ref{fig:splitting} (a)) and the $\Gamma$-point band splitting, thus
validating the low-energy description within the $\overline{\Gamma}_{10}$ subspace. Moreover, the mean-field decoupling of the interaction term in the AIAO phase, Eq. (\ref{eqn:micro-interaction}), reveals the origin of the $\Gamma$-point band splitting: because the electronic states behave as $J=3/2$ angular momentum states, it is possible to construct a momentum-independent combination of operators that transform as a magnetic octupolar moment.

To shed further light on this point, we go beyond our specific microscopic model and construct the most general low-energy Hamiltonian within the $\overline{\Gamma}_{10}$ subspace in the presence of a nonzero AIAO order parameter, which we denote for simplicity by $\Phi_{A_{2g}^-} $.   Following the procedure outlined in the beginning of this section, at each order in the momentum around the $\Gamma$ point, $\mathbf{k}$, we write down all independent Hermitian operators built from $J=3/2$ matrices and momentum polynomials that, when combined, transform as the $A_{2g}^-$ irreducible representation. 
 To quadratic order in $\mathbf{k}$, the resulting low-energy Hamiltonian is given by $\mathcal{H}_{\mathrm{eff}}=\mathcal{H}_{0}+\Phi_{A_{2g}^-} \mathcal{H}_{\mathrm{AIAO}}$, with $\mathcal{H}_{0}$ given in its most general form by Eq. (\ref{eq:H_hopping_projected}) and the coupling Hamiltonian by:
\begin{widetext}
        \begin{align}
            \mathcal{H}_{\mathrm{AIAO}}(\mathbf{k})=&\lambda_1 (k_y k_z J_x+k_x k_z J_y+k_x k_y J_z)+\lambda_2(k_y k_z J_x^3+k_x k_z J_y^3+k_x k_y J_z^3)+(\lambda_3+\lambda_4|\mathbf{k}|^2)(J_xJ_yJ_z+J_zJ_yJ_x), \label{eqn:symmetryA2g}
        \end{align}
\end{widetext}
where $\lambda_1$, $\lambda_2$, $\lambda_3$, and $\lambda_4$ are real-valued coupling constants between the electronic degrees of freedom and the AIAO order parameter  $\Phi_{A_{2g}^-} $. 
The explicit construction of this symmetry-allowed low-energy Hamiltonian in the $\overline\Gamma_{10}$ subspace, including its expansion up to quadratic order in momentum, is presented in Appendix~\ref{app:C}.

Comparison with the projected microscopic interaction in Eq.~\eqref{eqn:micro-interaction} identifies the $\lambda_3$ term in Eq.~\eqref{eqn:symmetryA2g} as the coupling realized in our mean-field decoupled microscopic model, with $\lambda_3=2\sqrt{3}US/3=0.0701$. Although this term carries no explicit $\mathbf{k}$-dependence, it produces a strongly momentum-dependent band splitting through its interplay with the hopping Hamiltonian [Fig.~\ref{fig:splitting}(d)]. In particular, it causes the band splitting at the $\Gamma$ point while, away from $\Gamma$, it leads to anisotropic band splitting through its combination with the non-interacting hopping Hamiltonian. 
 In contrast to the $\lambda_3$ term, the $\lambda_1$, $\lambda_2$, and $\lambda_4$ terms contain explicit $\mathbf{k}$-dependence and therefore only generate momentum-dependent band splittings in the vicinity of the $\Gamma$ point, but no splitting at $\Gamma$. To illustrate their effect, Fig.~\ref{fig:splitting}(b), (c), and (e) shows the band structures obtained by turning on each parameter separately. While all terms lift the Kramers degeneracy, the $\mathbf{k}$-dependent terms produce anisotropic splittings directly at low energies, even if one were to assume an isotropic hopping Hamiltonian.

It is interesting to compare the low-energy model in Eq. (\ref{eqn:symmetryA2g}) with the low-energy model for a one-band collinear $d$-wave altermagnet on the cubic lattice built from standard $S=1/2$ electronic states. To make the comparison more meaningful, we also include the effect of SOC in the altermagnetic model. The result, derived in Ref. \cite{Fernandes2024}, is:

\begin{equation}
 \mathcal{H}_{\mathrm{AM}}(\mathbf{k})=\xi(\mathbf{k})\sigma_0 + \tilde{\Phi}_{A_{2g}^-} (k_y k_z \sigma_x+k_x k_z \sigma_y+k_x k_y \sigma_z)
\end{equation}
where $\xi(\mathbf{k})$ is the dispersion in the non-magnetic phase, $\tilde{\Phi}_{A_{2g}^-}$ is the altermagnetic order parameter transforming as the $A_{2g}^-$ irrep, and $\sigma_i$ are Pauli matrices acting on spin space. Comparing it with  Eq. (\ref{eqn:symmetryA2g}), we see that the first term of the latter is equivalent to the spin-dependent and momentum-dependent term of the standard one-band altermagnetic low-energy model, with the spin $S=1/2$ operators replaced by the angular momentum $J=3/2$ dipolar operators. The additional terms in Eq. (\ref{eqn:symmetryA2g}) that are absent in the altermagnetic low-energy model are a consequence of the fact that certain cubic combinations of the $J_i$ operators transform either as a dipolar magnetic moment ($J_i^3$) or an octupolar magnetic moment ($J_xJ_yJ_z$). It is interesting that, despite the noncollinear, strongly spin-orbit-coupled character of the AIAO order, its effect on the low-energy electronic states can be captured in terms of a model that resembles a generalization of the $S=1/2$ altermagnetic model to $J=3/2$ states.
This also suggests that while the SOC is essential in forming the spin-orbit-coupled magnetic moments in this pyrochlore system, it is not solely responsible of the splitting of bands, which would have a non-relativistic component even if the underlying atomic moments originated from spins only. 

\section{Conclusion}

In this work, we established a microscopic and symmetry-based description of the AIAO state in pyrochlore iridates and demonstrated that it can be interpreted as a noncollinear counterpart of a $d$-wave altermagnetic phase stabilized by strong SOC. We showed that the AIAO order emerges at the mean-field level in a a realistic $j_{\mathrm{eff}}=1/2$ tight-binding model with onsite interaction, and that it transforms as an even-parity $A_{2g}^-$ magnetic octupole, breaking time-reversal symmetry while preserving a net zero magnetization.

Building on this microscopic model, we analyzed the consequences of the AIAO order for the electronic structure. Using group theory, we demonstrated that the AIAO order enforces a momentum-dependent lifting of band degeneracies, originating from the symmetry-driven decomposition of Kramers doublets. In contrast to other noncollinear magnets that display ferroic magnetic octupolar order that induce a nodeless spin-textured momentum-dependent band splitting in the $k_z=0$ plane, such as the $120^\circ$ order in the kagome lattice of Mn$_3$Sn, we showed that the AIAO order generates a nodal uniaxial spin-polarized band splitting at $k_z=0$, analogous to a collinear $d$-wave altermagnet. Moreover, the AIAO order also causes band splitting at the $\Gamma$-point and at the $X$-point (and symmetry-related momenta) by lowering the degeneracy of certain states from fourfold to twofold. Importantly, the residual twofold degeneracy corresponds to a pair of Kramers states, consistent with the absence of a net magnetization in the AIAO phase. In real-space, the momentum-dependent splitting is directly manifested in the multipolar decomposition of the magnetization density, which we calculated directly from our microscopic model.

To understand the origin of the band splittings, we constructed a low-energy effective Hamiltonian for the coupling between the fourfold-degenerate $\Gamma$-point states and the $A_{2g}^-$ AIAO order parameter, exploiting the standard parametrization of the former in terms of effective angular momentum $J=3/2$ degrees of freedom. We found that both the momentum-dependent and zero-momentum band splittings originate from a combination of the coupling between the AIAO order and the $J=3/2$ magnetic octupolar operator with the anisotropic hopping Hamiltonian containing $J=3/2$ electric quadrupolar operators. Comparing the symmetry-constructed model with a $\Gamma$-point expansion of our microscopic model, we identify the role of the microscopic Hubbard interaction in determining the coupling constants of the model.

Besides establishing the origin and character of the momentum-dependent band splittings caused by AIAO order, our work provides a roadmap to more direclty connect the properties of noncollinear spin-orbit coupled magnets with those of collinear altermagnets, despite the absence of a full formal spin-group classification of noncollinear phases. In this regard, it is interesting to note that, in the particular case of the pyrochlore iridates, any compensated collinear arrangement of the Ir magnetic moments would necessarily break the underlying cubic symmetry of the face-centered lattice, making it impossible to realize a cubic $d$-wave altermagnetic state in this structure. The SOC plays an essential role to ensure that the underlying cubic symmetry is preserved in the AIAO state, as it forces the moments to orient along the local $C_3$ axis determined by the octahedral environment.  As such, the AIAO order could be viewed as a ``collinear'' altermagnet in local reference frames, similar to how classical spin ice can be viewed as ferromagnetic in local frames. This analysis shows that SOC not only modifies the properties of collinear altermagnets, but it can also play a crucial role in enabling the emergence of certain altermagnetic-like properties. \\

\acknowledgments
We thank  Yong-Baek Kim, Johannes Knolle, Sergio di Matteo,  Joerg Schmallian, and Brad Ramshaw for fruitful discussions.
 N.B.P. were supported
by the U.S. Department of Energy, Office of Science, Basic Energy Sciences under Award No. DE-SC0018056. R.M.F.  acknowledges support from the Research Corporation for Science and Advancement through the Cottrell SEED Award CS-SEED-2025-012. T.B. was supported by the NSF CAREER grant DMR2046020.
~~\\

\appendix 

\section{Orbital and spin structure of the nearest-neighbor hopping in pyrochlore iridates}\label{app:A}

In this appendix we summarize the microscopic ingredients underlying
the nearest-neighbor hopping matrix $\hat{\mathcal{T}}^{ji}$ for the
Ir$^{4+}$ ions on the pyrochlore lattice and provide the technical
details omitted from the main text. We focus on the explicit
construction of the hopping amplitudes starting from the local
$t_{2g}$ orbital basis, including the effects of spin–orbit coupling,
local coordinate frames, and ligand-mediated processes.

We work in the local $t_{2g}$ orbital basis on each Ir site, where the orbital wave functions are defined with respect to site-dependent coordinate frames set by the surrounding IrO$_6$ octahedra. A convenient choice of local coordinates is the following:
\begin{align}
    &\begin{cases}
        \hat{x}_1=\left(\frac{2}{3},-\frac{1}{3},\frac{2}{3}\right)\\
        \hat{y}_1=\left(-\frac{1}{3},\frac{2}{3},\frac{2}{3}\right)\\
        \hat{z}_1=\left(-\frac{2}{3},-\frac{2}{3},\frac{1}{3}\right)
    \end{cases}, 
    \quad
    \begin{cases}
        \hat{x}_2=\left(\frac{2}{3},\frac{1}{3},-\frac{2}{3}\right)\\
        \hat{y}_2=\left(\frac{1}{3},\frac{2}{3},\frac{2}{3}\right)\\
        \hat{z}_2=\left(\frac{2}{3},-\frac{2}{3},\frac{1}{3}\right)
    \end{cases},\nonumber\\
    &\begin{cases}
        \hat{x}_3=\left(\frac{2}{3},\frac{1}{3},\frac{2}{3}\right)\\
        \hat{y}_3=\left(\frac{1}{3},\frac{2}{3},-\frac{2}{3}\right)\\
        \hat{z}_3=\left(-\frac{2}{3},\frac{2}{3},\frac{1}{3}\right)
    \end{cases},
    \quad
    \begin{cases}
        \hat{x}_4=\left(\frac{2}{3},-\frac{1}{3},-\frac{2}{3}\right)\\
        \hat{y}_4=\left(-\frac{1}{3},\frac{2}{3},-\frac{2}{3}\right)\\
        \hat{z}_4=\left(\frac{2}{3},\frac{2}{3},\frac{1}{3}\right)
    \end{cases}\nonumber,
\end{align}
where the coordinates of local axes are specified in the global coordinate. In Fig.~\ref{fig:1} (c), we let $(\hat{x}_1,\hat{y}_1,\hat{z}_1)=(\tilde{x},\tilde{y},\tilde{z})$ and $(\hat{x}_4,\hat{y}_4,\hat{z}_4)=(x',y',z')$. For a given
nearest-neighbor bond, the hopping matrix is obtained by explicitly
transforming the Slater-Koster overlap matrices \cite{SlaterKoster1954} between these local frames and the global crystallographic frame. Because the direct
$d$–$d$ and ligand-mediated $d$–$p$–$d$ processes are most naturally
formulated in different coordinate systems, they are combined only
after the appropriate basis transformations.

        \begin{widetext}
            \begin{center}
            \begin{table}[H]
                \centering
                \begin{tabular}{|c|c|c|c|c|c|}
                    \hline
                                   & $d_{yz}$ & $d_{zx}$ & $d_{xy}$ & $d_{3z^2-r^2}$ & $d_{x^2-y^2}$ \\\hline
                    $d_{yz}$       & $\frac{1}{2}dd\pi+\frac{1}{2}dd\delta$&$\frac{1}{2}dd\pi-\frac{1}{2}dd\delta$& $0$ & $0$& $0$\\\hline
                    $d_{zx}$       & $\frac{1}{2}dd\pi-\frac{1}{2}dd\delta$ & $\frac{1}{2}dd\pi+\frac{1}{2}dd\delta$ & $0$ & $0$ & $0$\\\hline
                    $d_{xy}$       & $0$ & $0$ & $\frac{3}{4}dd\sigma+\frac{1}{4}dd\delta$ & $-\frac{\sqrt{3}}{4}dd\sigma+\frac{\sqrt{3}}{4}dd\delta$ & $0$\\\hline
                    $d_{3z^2-r^2}$ & $0$  &$0$ & $-\frac{\sqrt{3}}{4}dd\sigma+\frac{\sqrt{3}}{4}dd\delta$ & $\frac{1}{4}dd\sigma+\frac{3}{4}dd\delta$ & $0$ \\\hline
                    $d_{x^2-y^2}$  & $0$  &$0$ & $0$ & $0$ & $dd\pi$ \\\hline
                \end{tabular}
                \caption{\label{tbl:sk}Slater-Koster parameters for $d$ orbitals on Ir${}_1$ and Ir${}_4$ ($l=1/\sqrt{2}$, $m=1/\sqrt{2}$, $n=0$) in global coordinates.}
            \end{table}
            \begin{table}[H]
                \centering
                \begin{tabular}{|c|c|c|c|}
                    \hline
                    & $d_{\tilde{y}\tilde{z}}(d_{y'z'})$ & $d_{\tilde{x}\tilde{z}}(d_{z'x'})$ & $d_{\tilde{x}\tilde{y}}(d_{x'y'})$\\\hline
                    $p_{\tilde{x}}(p_{x'}) $& $0$ & $pd\pi$ & $0$\\\hline
                    $p_{\tilde{y}}(p_{y'})$ & $pd\pi$ & $0$ & $0$\\\hline
                    $p_{\tilde{z}}(p_{z'})$ & $0$ & $0$ & $0$\\\hline
                \end{tabular}
                \caption{The Slater-Koster parameters for hopping between $p$ orbitals and $t_{2g}$ orbitals ($l=0$, $m=0$, $n=1$) in local coordinates.}
                \label{tbl:skpd}
            \end{table}
            \end{center}
        \end{widetext}

We therefore begin by determining the explicit transformation between
the $d$ orbitals defined in the local coordinate frames of each Ir ion
and those expressed in the global crystallographic frame. For the
representative Ir$_1$-Ir$_4$ bond shown in Fig.~\ref{fig:1} (c), this transformation corresponds to a
rotation about the $(1,-1,0)$ axis by an angle $\varphi$ for Ir$_1$ and
by $-\varphi$ for Ir$_4$, reflecting the opposite orientations of their
local octahedral environments. The resulting orbital basis
transformations for Ir$_1$ and Ir$_4$  are then given by
        \begin{widetext}
            \begin{align}
                d_{\tilde{y}\tilde{z}}&=\frac{\cos2\varphi+\cos\varphi}{2}d_{yz}+\frac{\cos2\varphi-\cos\varphi}{2}d_{zx}-\frac{\sin2\varphi}{2\sqrt{2}}d_{xy}+\frac{\sqrt{3}\sin2\varphi}{2\sqrt{2}} d_{3z^2-r^2}+\frac{\sin\varphi}{\sqrt{2}}d_{x^2-y^2},\nonumber\\
                d_{\tilde{z}\tilde{x}}&=\frac{\cos2\varphi-\cos\varphi}{2}d_{yz}+\frac{\cos2\varphi+\cos\varphi}{2}d_{zx}-\frac{\sin2\varphi}{2\sqrt{2}}d_{xy}+\frac{\sqrt{3}\sin2\varphi}{2\sqrt{2}} d_{3z^2-r^2}-\frac{\sin\varphi}{\sqrt{2}}d_{x^2-y^2},\\
                d_{\tilde{x}\tilde{y}}&=\frac{\sin2\varphi}{2\sqrt{2}}d_{yz}+\frac{\sin2\varphi}{2\sqrt{2}}d_{zx}+\frac{3+\cos2\varphi}{4}d_{xy}+\frac{\sqrt{3}(1-\cos2\varphi)}{4}d_{3z^2-r^2}, \nonumber
            \end{align}
            and for Ir${}_4$:
            \begin{align}
                d_{y'z'}&=\frac{\cos2\varphi+\cos\varphi}{2}d_{yz}+\frac{\cos2\varphi-\cos\varphi}{2}d_{zx}+\frac{\sin2\varphi}{2\sqrt{2}}d_{xy}-\frac{\sqrt{3}\sin2\varphi}{2\sqrt{2}} d_{3z^2-r^2}-\frac{\sin\varphi}{\sqrt{2}}d_{x^2-y^2},\nonumber\\
                d_{z'x'}&=\frac{\cos2\varphi-\cos\varphi}{2}d_{yz}+\frac{\cos2\varphi+\cos\varphi}{2}d_{zx}+\frac{\sin2\varphi}{2\sqrt{2}}d_{xy}-\frac{\sqrt{3}\sin2\varphi}{2\sqrt{2}} d_{3z^2-r^2}+\frac{\sin\varphi}{\sqrt{2}}d_{x^2-y^2},\\
                d_{x'y'}&=-\frac{\sin2\varphi}{2\sqrt{2}}d_{yz}-\frac{\sin2\varphi}{2\sqrt{2}}d_{zx}+\frac{3+\cos2\varphi}{4}d_{xy}+\frac{\sqrt{3}(1-\cos2\varphi)}{4}d_{3z^2-r^2}, \nonumber
            \end{align}
        \end{widetext}
        
        For the ideal pyrochlore geometry, the relative orientation of the local octahedral environments uniquely fixes the rotation angle to $\varphi = \arctan(2\sqrt{2})$. With this choice, the orbital basis transformations on Ir$_1$ and  Ir$_4$
         take the explicit matrix forms:
        \begin{align}
            \begin{pmatrix}
                d_{\tilde{y}\tilde{z}}\\
                d_{\tilde{z}\tilde{x}}\\
                d_{\tilde{x}\tilde{y}}
            \end{pmatrix}=
            \begin{pmatrix}
                -\frac{2}{9}&-\frac{5}{9}&-\frac{2}{9}&\frac{2}{3\sqrt{3}}&\frac{2}{3}\\
                -\frac{5}{9}&-\frac{2}{9}&-\frac{2}{9}&\frac{2}{3\sqrt{3}}&-\frac{2}{3}\\
                \frac{2}{9}&\frac{2}{9}&\frac{5}{9}&\frac{4}{3\sqrt{3}}&0
            \end{pmatrix}
            \begin{pmatrix}
                d_{yz}\\
                d_{zx}\\
                d_{zx}\\
                d_{3z^2-r^2}\\
                d_{x^2-y^2}
            \end{pmatrix}\label{eqn:Ir1glb2loc}
        \end{align}
        and
        \begin{align}
            \begin{pmatrix}
                d_{y'z'}\\
                d_{z'x'}\\
                d_{x'y'}
            \end{pmatrix}=
            \begin{pmatrix}
                -\frac{2}{9}&-\frac{5}{9}&\frac{2}{9}&-\frac{2}{3\sqrt{3}}&-\frac{2}{3}\\
                -\frac{5}{9}&-\frac{2}{9}&\frac{2}{9}&-\frac{2}{3\sqrt{3}}&\frac{2}{3}\\
                -\frac{2}{9}&-\frac{2}{9}&\frac{5}{9}&\frac{4}{3\sqrt{3}}&0
            \end{pmatrix}
            \begin{pmatrix}
                d_{yz}\\
                d_{zx}\\
                d_{zx}\\
                d_{3z^2-r^2}\\
                d_{x^2-y^2}
            \end{pmatrix}.\label{eqn:Ir4glb2loc}
        \end{align}
       We denote the resulting $3\times5$ matrices in Eqs.~\eqref{eqn:Ir1glb2loc} and \eqref{eqn:Ir4glb2loc}
         by $\mathcal{L}^{\mathrm{dd}}_1$ and $\mathcal{L}^{\mathrm{dd}}_4$, respectively. 
Using this notation, the direct hopping between the local $t_{2g}$ orbitals on
Ir$_1$ and Ir$_4$ can be written as
      \begin{align}
            D'^\dagger \tilde{\mathcal{T}}^{14}_{\mathrm{direct}} \tilde{D}
            \equiv 
            D'^\dagger(\mathcal{L}_4^{\mathrm{dd}} \mathcal{M}_{14}^{\mathrm{dd}} \mathcal{L}_1^{\mathrm{dd}\dagger}) \tilde{D},
        \end{align}
        where $D'^\dagger= (d_{y'z'}^\dagger \quad d_{z'x'}^\dagger \quad d_{x'y'}^\dagger)$ and  $\tilde{D} = (d_{\tilde{y}\tilde{z}} \quad d_{\tilde{z}\tilde{x}} \quad d_{\tilde{x}\tilde{y}})^T$ are vectors of local $t_{2g}$ operators on Ir$_4$ and Ir$_1$, respectively. The matrix $\mathcal{M}^{\mathrm{dd}}_{14}$ is the $5\times5$ Slater-Koster matrix obtained by rewriting Table~\ref{tbl:sk} in matrix form, where Ir$_4$ and Ir$_1$ are connected by the unit vector $(l,m,n)=(1,1,0)/\sqrt{2}$ in the global coordinates. 
 
 The resulting effective hopping matrix $\tilde{\mathcal{T}}_{\mathrm{direct}}^{14}$ between local $t_{2g}$ orbitals takes the form        \begin{align}\label{eqn:Tdirect14}
            \tilde{\mathcal{T}}^{14}_{\mathrm{direct}}=
            \begin{pmatrix}
                t_1 & t_2 & -t_4\\
                t_2 & t_1 & -t_4\\
                t_4 & t_4 & t_3
            \end{pmatrix},
        \end{align}
        with the matrix elements given by
        \begin{align}
            t_1&=\frac{1}{162}(-24dd\sigma-23dd\pi+dd\delta),\nonumber\\
            t_2&=\frac{1}{162}(-24dd\sigma+121dd\pi-17dd\delta),\nonumber\\
            t_3&=\frac{1}{324}(3dd\sigma-32dd\pi+289dd\delta),\\
            t_4&=\frac{1}{81}(-3dd\sigma+14dd\pi+17dd\delta).\nonumber
        \end{align}
        Analogous expressions can be derived for all other bonds within the tetrahedron, yielding the remaining hopping matrices:
        \begin{align}
            &\tilde{\mathcal{T}}^{12}_{\mathrm{direct}}=\begin{pmatrix}
                -t_1 &-t_4 & t_2\\
                 t_4 &-t_3 &-t_4\\
                -t_2 &-t_4 & t_1
            \end{pmatrix},\nonumber\\
            &\tilde{\mathcal{T}}^{13}_{\mathrm{direct}}=\begin{pmatrix}
                -t_3 & t_4 &-t_4\\
                -t_4 &-t_1 & t_2\\
                -t_4 &-t_2 & t_1
            \end{pmatrix},\nonumber\\
            &\tilde{\mathcal{T}}^{23}_{\mathrm{direct}}=\begin{pmatrix}
                 t_1 &-t_2 & t_4\\
                -t_2 & t_1 &-t_4\\
                -t_4 & t_4 & t_3
            \end{pmatrix},\\
            &\tilde{\mathcal{T}}^{24}_{\mathrm{direct}}=\begin{pmatrix}
                -t_3 &-t_4 & t_4\\
                 t_4 &-t_1 & t_2\\
                 t_4 &-t_2 & t_1
            \end{pmatrix},\nonumber\\
            &\tilde{\mathcal{T}}^{34}_{\mathrm{direct}}=\begin{pmatrix}
                -t_1 & t_4 & t_2\\
                -t_4 &-t_3 & t_4\\
                -t_2 & t_4 & t_1
            \end{pmatrix}.\nonumber
        \end{align}

        To evaluate the ligand-mediated hopping contributions, we first determine the transformation between the local coordinate systems of the ligand $p$ orbitals and the global crystallographic frame. Each ligand sits between two neighboring Ir ions.
        Ir$_1$ and Ir$_4$ and its $p$ orbitals  are most naturally defined in local coordinate frames aligned with the corresponding Ir–ligand bond  directions.

         For the ligand adjacent to Ir$_1$, the transformation from the local $(\tilde{x},\tilde{y},\tilde{z})$ frame to the global $(x,y,z)$ frame (with local axes defined in Fig.~\ref{fig:1} (c)) is
        \begin{align}
            p_x=&\cos^2\frac{\varphi}{2}p_{\tilde{x}}+\frac{\cos\varphi-1}{2}p_{\tilde{y}}-\frac{\sin\varphi}{\sqrt{2}}p_{\tilde{z}},\nonumber\\
            p_y=&\frac{\cos\varphi-1}{2}p_{\tilde{x}}+\cos^2\frac{\varphi}{2}p_{\tilde{y}}-\frac{\sin\varphi}{\sqrt{2}}p_{\tilde{z}},\\
            p_z=&\frac{\sin\varphi}{\sqrt{2}}p_{\tilde{x}}+\frac{\sin\varphi}{\sqrt{2}}p_{\tilde{y}}+\cos\varphi p_{\tilde{z}}.\nonumber
        \end{align}
       Similarly, for the ligand adjacent to Ir$_4$, the transformation from the local $(x',y',z')$ frame reads
        \begin{align}
            p_x=&\cos^2\frac{\varphi}{2}p_{x'}+\frac{\cos\varphi-1}{2}p_{y'}+\frac{\sin\varphi}{\sqrt{2}}p_{z'},\nonumber\\
            p_y=&\frac{\cos\varphi-1}{2}p_{x'}+\cos^2\frac{\varphi}{2}p_{y'}+\frac{\sin\varphi}{\sqrt{2}}p_{z'},\\
            p_z=&-\frac{\sin\varphi}{\sqrt{2}}p_{x'}-\frac{\sin\varphi}{\sqrt{2}}p_{y'}+\cos\varphi p_{z'}.\nonumber
        \end{align}
       For the ideal pyrochlore geometry with the rotation angle $\varphi=\arctan 2\sqrt{2}$, we get
        \begin{align}
            \begin{pmatrix}
                p_x\\
                p_y\\
                p_z
            \end{pmatrix}=
            \begin{pmatrix}
                 \frac{2}{3}&-\frac{1}{3}&-\frac{2}{3}\\
                -\frac{1}{3}&\frac{2}{3}&-\frac{2}{3}\\
                 \frac{2}{3}&\frac{2}{3}&\frac{1}{3}
            \end{pmatrix}
            \begin{pmatrix}
                p_{\tilde{x}}\\
                p_{\tilde{y}}\\
                p_{\tilde{z}}
            \end{pmatrix}\label{eqn:pp1}
        \end{align}
        and
        \begin{align}
            \begin{pmatrix}
                p_x\\
                p_y\\
                p_z
            \end{pmatrix}=
            \begin{pmatrix}
                 \frac{2}{3}&-\frac{1}{3}&\frac{2}{3}\\
                -\frac{1}{3}&\frac{2}{3}&\frac{2}{3}\\
                -\frac{2}{3}&-\frac{2}{3}&\frac{1}{3}
            \end{pmatrix}
            \begin{pmatrix}
                p_{x'}\\
                p_{y'}\\
                p_{z'}
            \end{pmatrix}\label{eqn:pp4}.
        \end{align}
We denote the $3\times3$ transformation matrices appearing in
Eqs.~\eqref{eqn:pp1} and \eqref{eqn:pp4} by $\mathcal{L}^{\mathrm{pp}}_1$ and
$\mathcal{L}^{\mathrm{pp}}_4$, respectively. Using these matrices, the ligand-mediated hopping between Ir$_1$ and Ir$_4$ can be
derived within second-order perturbation theory in the $d$-$p$ hybridization.
Explicitly, the effective hopping between local $t_{2g}$ orbitals $|d_\alpha\rangle$
on Ir$_4$ and $|d_\beta\rangle$ on Ir$_1$ is then  given by
        \begin{eqnarray}
            &&\tilde{\mathcal{T}}^{14}_{\mathrm{med},\alpha\beta}| d_\alpha\rangle\langle d_\beta|=\\\nonumber && \sum_{\mu\nu}\sum_{\gamma}^{xyz}\frac{\langle d_\alpha|\mathcal{M}_{\alpha \mu}^{\mathrm{pd}\dagger}\mathcal{L}_{4,\mu \gamma}^{\mathrm{pp}\dagger}|p_\gamma\rangle\langle p_\gamma|\mathcal{L}_{1,\gamma \nu}^{\mathrm{pp}}\mathcal{M}_{\nu \beta}^{\mathrm{pd}}|d_{\beta}\rangle }{\Delta_{\mathrm{pd}}}| d_\alpha\rangle\langle d_\beta|,
        \end{eqnarray}
where $\alpha$ and $\beta$ label the local $t_{2g}$ orbitals, $\mu=x',y',z'$ and $\nu=\tilde{x},\tilde{y},\tilde{z}$
refer to local ligand orbitals, and
$\gamma=x,y,z$ labels the global $p$ orbitals. The energy
$\Delta_{\mathrm{pd}}$ denotes the charge-transfer gap between the Ir $d$ states
and the ligand $p$ states, and $\mathcal{M}^{\mathrm{pd}}$ is the $3\times3$
Slater-Koster hopping matrix listed in Table~\ref{tbl:skpd}, where Ir and O ions are always connected by the unit vector $(l,m,n)=(0,0,1)$ in the local coordinates.
 This gives the effective ligand-mediated hopping matrix
        \begin{align}
            \tilde{\mathcal{T}}^{14}_{\mathrm{med}}=\frac{pd\pi^2}{\Delta_{\mathrm{pd}}}
            \begin{pmatrix}
            \frac{1}{9} & -\frac{8}{9} & 0\\
            -\frac{8}{9}& \frac{1}{9} &0\\
            0 & 0 & 0
            \end{pmatrix}.
        \end{align}
        
Comparing this result with the direct hopping matrix in
Eq.~\eqref{eqn:Tdirect14}, we observe that the ligand-mediated contribution preserves
the overall matrix structure and leads only to a renormalization of the parameters
$t_1$ and $t_2$.
The total local hopping matrix $\tilde{\mathcal{T}}^{14}_{\mathrm{orbital}}$ 
therefore retains the same form as
Eq.~\eqref{eqn:Tdirect14}, with the replacements 
        \begin{align}
            t_1'&=\frac{1}{162}(-24dd\sigma-23dd\pi+dd\delta) + \frac{1}{9}\frac{pd\pi^2}{\Delta_{\mathrm{pd}}},\nonumber\\
            t_2'&=\frac{1}{162}(-24dd\sigma+121dd\pi-17dd\delta)- \frac{8}{9}\frac{pd\pi^2}{\Delta_{\mathrm{pd}}}.
        \end{align}
        
        Having determined the orbital part of the hopping matrix between Ir$_1$ and Ir$_4$, 
        we now turn to the spin part. The $j_{\mathrm{eff}}=1/2$ pseudospin degrees of freedom are defined in site-dependent local quantization frames set by the orientations of the
IrO$_6$ octahedra. As a result, hopping between Ir$_1$ and Ir$_4$ involves not only orbital overlap but also a rotation that relates the corresponding local spin frames. This rotation is described by the $SU(2)$ Wigner rotation matrices $\mathcal{D}_1$ and $\mathcal{D}_4$, which rotate the global spin quantization axis into the local axes of Ir$_1$ and Ir$_4$, respectively. The spinor part of the hopping on the Ir$_1$–Ir$_4$
bond is therefore given by
        \begin{align}
            \begin{pmatrix}
                d_{\uparrow'}^\dagger & d_{\downarrow'}^\dagger
            \end{pmatrix}
            \tilde{\mathcal{T}}^{14}_{\mathrm{spin}}\begin{pmatrix}
                d_{\tilde{\uparrow}}\\d_{\tilde{\downarrow}}
            \end{pmatrix}=
            \begin{pmatrix}
                d_{\uparrow'}^\dagger & d_{\downarrow'}^\dagger
            \end{pmatrix}
            \mathcal{D}_4^\dagger \mathcal{D}_1
            \begin{pmatrix}
                d_{\tilde{\uparrow}}\\d_{\tilde{\downarrow}}
            \end{pmatrix}.
        \end{align}
        Using the local axes, the $SU(2)$ rotation matrices for
the four Ir sublattices take the form
        \begin{align}
            \mathcal{D}_1=\begin{pmatrix}
                \sqrt{\frac{2}{3}} & \frac{1-i}{\sqrt{6}}\\
                \frac{-1-i}{\sqrt{6}} & \sqrt{\frac{2}{3}}
            \end{pmatrix},\,\mathcal{D}_2=\begin{pmatrix}
                \sqrt{\frac{2}{3}} & \frac{-1-i}{\sqrt{6}}\\
                \frac{1-i}{\sqrt{6}} & \sqrt{\frac{2}{3}}
            \end{pmatrix},\,\nonumber\\
            \mathcal{D}_3=\begin{pmatrix}
                \sqrt{\frac{2}{3}} & \frac{1+i}{\sqrt{6}}\\
                \frac{-1+i}{\sqrt{6}} & \sqrt{\frac{2}{3}}
            \end{pmatrix},\,
            \mathcal{D}_4=\begin{pmatrix}
                \sqrt{\frac{2}{3}} & \frac{-1+i}{\sqrt{6}}\\
                \frac{1+i}{\sqrt{6}} & \sqrt{\frac{2}{3}}
            \end{pmatrix}.
        \end{align}
        The resulting spinor hopping matrix between Ir$_1$ and Ir$_4$ is
        \begin{align}
            \tilde{\mathcal{T}}_{\mathrm{spin}}^{14}=
            \begin{pmatrix}
                \frac{1}{3} & \frac{2}{3}-\frac{2}{3}i\\
                -\frac{2}{3}-\frac{2}{3}i&\frac{1}{3}
            \end{pmatrix}.
        \end{align}
        The spinor hopping matrices on the remaining bonds are obtained analogously as
$\tilde{\mathcal{T}}^{ji}_{\mathrm{spin}} = \mathcal{D}_j^\dagger \mathcal{D}_i$.
For completeness, we list them below:
\begin{eqnarray}
            &&\tilde{\mathcal{T}}_{\mathrm{spin}}^{12}=\begin{pmatrix}
                \frac{2}{3}-\frac{1}{3}i &\frac{2}{3}\\
                -\frac{2}{3} &\frac{2}{3}+\frac{1}{3}i
            \end{pmatrix},\nonumber\\&&
            \tilde{\mathcal{T}}_{\mathrm{spin}}^{13}=\begin{pmatrix}
                \frac{2}{3}+\frac{1}{3}i &-\frac{2}{3}i\\
                -\frac{2}{3}i &\frac{2}{3}-\frac{1}{3}i
            \end{pmatrix},\nonumber\\&&
            \tilde{\mathcal{T}}_{\mathrm{spin}}^{23}=\begin{pmatrix}
                \frac{1}{3} &-\frac{2}{3}-\frac{2}{3}i\\
                \frac{2}{3}-\frac{2}{3}i&\frac{1}{3}
            \end{pmatrix},\\&&
            \tilde{\mathcal{T}}_{\mathrm{spin}}^{24}=\begin{pmatrix}
                \frac{2}{3}-\frac{1}{3}i &-\frac{2}{3}i\\
                -\frac{2}{3}i &\frac{2}{3}+\frac{1}{3}i
            \end{pmatrix},\nonumber\\&&\nonumber
            \tilde{\mathcal{T}}_{\mathrm{spin}}^{34}=\begin{pmatrix}
                \frac{2}{3}+\frac{1}{3}i &\frac{2}{3}\\
                -\frac{2}{3} &\frac{2}{3}-\frac{1}{3}i
            \end{pmatrix}.
        \end{eqnarray}

        The full hopping matrix between two local $t_{2g}$ orbitals is formed by the tensor product of the orbital part and the spinor part,
        \begin{eqnarray}
            \tilde{\mathcal{T}}_{\mathrm{total}}=\tilde{\mathcal{T}}_{\mathrm{orbital}}\otimes \tilde{\mathcal{T}}_{\mathrm{spin}}.
        \end{eqnarray}
        To obtain the effective hopping between two global $j_{\mathrm{eff}} = 1/2$ states, we project $\tilde{\mathcal{T}}_{\mathrm{total}}$ between two local $j_{\mathrm{eff}} = 1/2$ using the basis transformation matrix $O_J$ 
        \begin{align}
            O_{J}=
            \begin{pmatrix}
                0 & \tfrac{1}{\sqrt{3}} & 0 & \tfrac{i}{\sqrt{3}} & \tfrac{1}{\sqrt{3}} & 0\\
                \tfrac{1}{\sqrt{3}} & 0 & -\tfrac{i}{\sqrt{3}} & 0 & 0 & -\tfrac{1}{\sqrt{3}}\\
                -\tfrac{1}{\sqrt{2}} & 0 & -\tfrac{i}{\sqrt{2}} & 0 & 0 & 0\\
                0 & -\tfrac{1}{\sqrt{6}} & 0 & -\tfrac{i}{\sqrt{6}} & \tfrac{2}{\sqrt{6}} & 0\\
                \tfrac{1}{\sqrt{6}} & 0 & -\tfrac{i}{\sqrt{6}} & 0 & 0 & \tfrac{2}{\sqrt{6}}\\
                0 & \tfrac{1}{\sqrt{2}} & 0 & -\tfrac{i}{\sqrt{2}} & 0 & 0
           \end{pmatrix},
           \label{eqn:O2J-matrix}
        \end{align}
        which transforms the orbital basis to the $j_{\mathrm{eff}}$ basis through
        \begin{align}
            \begin{pmatrix}
              d^\dagger_{1/2,+1/2}\\
              d^\dagger_{1/2,-1/2}\\
              d^\dagger_{3/2,+3/2}\\
              d^\dagger_{3/2,+1/2}\\
              d^\dagger_{3/2,-1/2}\\
            d^\dagger_{3/2,-3/2}
            \end{pmatrix}=O_{J}\,\begin{pmatrix}
              d^\dagger_{yz,\uparrow}\\
              d^\dagger_{yz,\downarrow}\\
              d^\dagger_{zx,\uparrow}\\
              d^\dagger_{zx,\downarrow}\\
              d^\dagger_{xy,\uparrow}\\
              d^\dagger_{xy,\downarrow}
            \end{pmatrix}.
        \end{align}
        On the Ir$_1$-Ir$_4$ bond, taking the upper left $2\times2$ block of $O_J^*\mathcal{T}^{14}_{\mathrm{total}}O_J^T$ yields
        \begin{align}
            \tilde{\mathcal{T}}^{14}_{J}&=
            \begin{pmatrix}
                [O_J^*\tilde{\mathcal{T}}^{14}_{\mathrm{total}}O_J^T]_{11} & [O_J^*\tilde{T}^{14}_{\mathrm{total}}O_J^T]_{12}\\
                [O_J^*{\mathcal{T}}^{14}_{\mathrm{total}}O_J^T]_{21} & [O_J^*{\mathcal{T}}^{14}_{\mathrm{total}}O_J^T]_{22}
            \end{pmatrix}\\\nonumber
            &= \frac{1}{9}
            \begin{pmatrix}
               2t_1'+t_3+8t_4 & 2(-1+i)(2t_2'+t_3-t_4)\\
                2(1+i)(2t_2'+t_3-t_4) & 2t_1'+t_3+8t_4
            \end{pmatrix}.
        \end{align}
        Similarly, we can also obtain the remaining projected hopping matrices:
        \begin{widetext}
            \begin{align}
                \tilde{\mathcal{T}}^{12}_J &= \frac{1}{9}\begin{pmatrix}
                    -(2i t_1'+4t_2'+(2+i)t_3-(2-8i)t_4) &-2(2t_2'+t_3-t_4)\\
                    2(2t_2'+t_3-t_4) & 2it_1'-4t_2'-(2-i)t_3+(2+8i)t_4
                \end{pmatrix},\nonumber\\
                \tilde{\mathcal{T}}^{13}_J &=  \frac{1}{9}\begin{pmatrix}
                   2i t_1'-4t_2'-(2-i)t_3+(2+8i)t_4 &2i(2t_2'+t_3-t_4)\\
                    2i(2t_2'+t_3-t_4) &-(2i t_1'+4t_2'+(2+i)t_3-(2-8i)t_4)
                \end{pmatrix},\nonumber\\
                \tilde{\mathcal{T}}^{23}_J &= \frac{1}{9}\begin{pmatrix}
                    2t_1'+t_3+8t_4 & 2(1+i)(2t_2'+t_3-t_4)\\
                    2(-1+i)(2t_2'+t_3-t_4) & 2t_1'+t_3+8t_4
                \end{pmatrix},\\
                \tilde{\mathcal{T}}^{24}_J &= \frac{1}{9}\begin{pmatrix}
                    -(2i t_1'+4t_2'+(2+i)t_3-(2-8i)t_4)&2i(2t_2'+t_3-t_4)\\
                    2i(2t_2'+t_3-t_4) & 2i t_1'-4t_2'-(2-i)t_3+(2+8i)t_4 
                \end{pmatrix},\nonumber\\
                \tilde{\mathcal{T}}^{34}_J &= \frac{1}{9}\begin{pmatrix}
                    2it_1'-4t_2'-(2-i)t_3+(2+8i)t_4&-2(2t_2'+t_3-t_4)\\
                    2(2t_2'+t_3-t_4) &-(2i t_1'+4t_2'+(2+i)t_3-(2-8i)t_4)
                \end{pmatrix}.\nonumber
            \end{align}
        \end{widetext}
   
 Finally, the effective hopping between two global
$j_{\mathrm{eff}}=1/2$ states is obtained by rotating the projected
local hopping matrices back into the global pseudospin
frame using the site-dependent $SU(2)$ rotation matrices
$\mathcal{D}_i$ introduced above,
$\mathcal{T}^{ji} = \mathcal{D}_i \tilde{\mathcal{T}}^{ji}_{J}
    \mathcal{D}_j^\dagger $.
Carrying out this transformation explicitly for all
nearest-neighbor bonds yields a hopping matrix whose
spin structure is fully constrained by lattice symmetry
and spin–orbit coupling. The resulting expression takes
the compact form given in Eq.~(\ref{eqn:hoppingGlobal})
of the main text:
        \begin{align}\label{eqn:hoppingGlobal-app}
            \mathcal{T}^{ji}=t\, \mathbb{I}_2+ i t'\, \mathbf{d}_{ji} \cdot \boldsymbol{\sigma},
        \end{align}
        where $\mathbb{I}_2$ is the $2\times 2$ identity, the vectors $\mathbf{d}_{ij}$ coincide with the pyrochlore's Dzyaloshinskii–Moriya axes,
        \begin{align}
            \mathbf{d}_{12} = (0,-1,1),\quad\mathbf{d}_{13}=(1,0,-1),\quad \mathbf{d}_{14} = (-1,1,0),\nonumber\\
            \mathbf{d}_{23} = (-1, -1,0),\quad\mathbf{d}_{24}=(1,0,1),\quad \mathbf{d}_{34} = (0,-1,-1),
        \end{align}
        $\boldsymbol{\sigma}$ denotes Pauli matrices, and
        \begin{align}
            t&=\frac{1}{27}(2t_1'-16t_2'-7t_3+16t_4)\nonumber\\
            &=\frac{1}{972}\left(51dd\sigma-316dd\pi-43dd\delta+520\frac{pd\pi^2}{\Delta_{pd}}\right),\\
            t'&=-\frac{2}{27}[2(t_1'+t_2'+t_3)+7t_4]\nonumber\\
            &=\frac{1}{972}\left(60dd\sigma-160dd\pi-220dd\delta+112\frac{pd\pi^2}{\Delta_{pd}}\right).
        \end{align} 
        
      The coefficients entering this final form encode the
combined effects of direct $d$–$d$ overlap and
ligand-mediated $d$–$p$–$d$ processes, while the
bond-dependent vectors reflect the geometry of the
pyrochlore lattice. This completes the microscopic
derivation of the nearest-neighbor hopping Hamiltonian
used in the main text.  

\section{Projection onto the $\overline{\Gamma}_{10}$ subspace}
    \label{app:B}
    
    Around the $\Gamma$ point, the microscopic Hamiltonian can be projected onto the four-dimensional $\overline{\Gamma}_{10}$ subspace, where it can be expressed in terms of effective angular-momentum $3/2$ matrices $J_x$, $J_y$ $J_z$ \cite{Luttinger1955,Luttinger1956}.

    We define a basis in the $\overline{\Gamma}_{10}$ subspace in terms of the
 $|J,J_z\rangle$ states as
    \begin{align}
        |v_1\rangle&=\frac{1}{\sqrt{2}}\left|\frac{3}{2},+\frac{1}{2}\right\rangle-\frac{i}{\sqrt{2}}\left|\frac{3}{2},-\frac{3}{2}\right\rangle,\nonumber\\
        |v_2\rangle&=\frac{i}{\sqrt{2}}\left|\frac{3}{2},+\frac{3}{2}\right\rangle-\frac{1}{\sqrt{2}}\left|\frac{3}{2},-\frac{1}{2}\right\rangle,\nonumber\\
        |v_3\rangle&=\left(-\frac{i}{\sqrt{2}}\left|\frac{3}{2},+\frac{3}{2}\right\rangle-\frac{1}{\sqrt{2}}\left|\frac{3}{2},-\frac{1}{2}\right\rangle\right)e^{-i\pi/4},\label{eqn:3halfbasis}\\
        |v_4\rangle&=\left(\frac{1}{\sqrt{2}}\left|\frac{3}{2},+\frac{1}{2}\right\rangle+\frac{i}{\sqrt{2}}\left|\frac{3}{2},-\frac{3}{2}\right\rangle\right)e^{i 5\pi/4}.\nonumber
    \end{align}
following the convention of Altmann \& Herzig \cite{Altmann1994}. In this basis, the $\overline{\Gamma}_{10}$ subspace realizes the standard spin-$3/2$ representation, so that the projected Hamiltonian can be expressed in terms of the operators $J_x$, $J_y$, and $J_z$. The corresponding matrix representations are
    \begin{align}\label{Jmatrices}
        J_x&=\begin{pmatrix}
            0 & -\frac{1}{2} & e^{-i11\pi/12} & 0\\
            -\frac{1}{2} & 0 & 0 & e^{i7\pi/12}\\
            e^{i11\pi/12} & 0 & 0 & \frac{i}{2}\\
            0 & e^{-i7\pi/12} & -\frac{i}{2} & 0
        \end{pmatrix},\nonumber\\
        J_y&=\begin{pmatrix}
            0 & \frac{i}{2} & e^{-i\pi/12} & 0\\
            -\frac{i}{2} & 0 & 0 & e^{i5\pi/12}\\
            e^{i\pi/12} & 0 & 0 &-\frac{1}{2}\\
            0 & e^{-i5\pi/12} & -\frac{1}{2} & 0
        \end{pmatrix},\\
        J_z&=\begin{pmatrix}
            -\frac{1}{2} & 0 & 0 & e^{-i3\pi/4}\\
            0 & \frac{1}{2} & e^{i3\pi/4} & 0\\
            0 & e^{-i3\pi/4} & \frac{1}{2} & 0\\
            e^{i3\pi/4} & 0 & 0 & -\frac{1}{2}
        \end{pmatrix}.\nonumber
    \end{align}
 
   These matrices satisfy the usual angular momentum algebra $[J_\alpha,J_\beta]=i\epsilon_{\alpha\beta\gamma}J_\gamma$ and thus provide a faithful realization of the $J=3/2$ representation. In what follows, we use this representation to project the microscopic Hamiltonian onto the $\overline{\Gamma}_{10}$ subspace.

    \subsection{Hopping term}
    We first project the hopping part of the microscopic Hamiltonian \eqref{eqn:tightbinding} onto $\overline{\Gamma}_{10}$. Inversion symmetry restricts the $\Gamma$-point expansion to even orders of $\mathbf{k}$, so we consider the expansion up to the quadratic order of $\mathbf{k}$. At the quadratic level, the bilinears of $\mathbf{k}$ transform as $A_{1g}$: $k_x^2+k_y^2+k_z^2$, $E_g$: ($\sqrt{3}(k_x^2-k_y^2)$, $2k_z^2-k_x^2-k_y^2$), and $T_{2g}$: ($k_y k_z$, $k_x k_z$, $k_x k_y$). Since the hopping Hamiltonian preserves the time reversal symmetry, the projected Hamiltonian must couple with $A_{1g}$, $E_g$, and $T_{2g}$ bilinears of $J_x$, $J_y$, and $J_z$, which are
    \begin{align}
        A_{1g}: \quad&  \frac{4}{15}(J_x^2+J_y^2+J_z^2),\nonumber\\
        E_{g}:  \quad&  \frac{1}{\sqrt{3}}(J_x^2-J_y^2),\nonumber\\
                \quad&  \frac{1}{3}(2J_z^2-J_x^2-J_y^2),\nonumber\\
        T_{2g}: \quad&  \frac{1}{\sqrt{3}}(J_yJ_z+J_zJ_y),\label{eqn:J2channels}\\
                \quad&  \frac{1}{\sqrt{3}}(J_xJ_z+J_zJ_x),\nonumber\\
                \quad&  \frac{1}{\sqrt{3}}(J_xJ_y+J_yJ_x).\nonumber
    \end{align}

To expand the hopping Hamiltonian within the $\overline{\Gamma}_{10}$ subspace, we first choose a basis  in which the Hamiltonian is block diagonal at the $\Gamma$ point with respect to the $\overline{\Gamma}_6$, $\overline{\Gamma}_7$, and $\overline{\Gamma}_{10}$ sectors.  
We work in this basis throughout, so that the projected Hamiltonian is expressed directly in terms of the spin-$3/2$ operators $J_x$, $J_y$, and $J_z$.
The corresponding basis transformation is given by
\begin{widetext}
    \begin{align}\label{eqn:projectU}
        \scalebox{0.9}{$
        \begin{pmatrix}
            |\psi^{\overline{\Gamma}_6}_1\rangle\\
            |\psi^{\overline{\Gamma}_6}_2\rangle\\
            |\psi^{\overline{\Gamma}_7}_1\rangle\\
            |\psi^{\overline{\Gamma}_7}_2\rangle\\
            |\psi^{\overline{\Gamma}_{10}}_1\rangle\\
            |\psi^{\overline{\Gamma}_{10}}_2\rangle\\
            |\psi^{\overline{\Gamma}_{10}}_3\rangle\\
            |\psi^{\overline{\Gamma}_{10}}_4\rangle\\
        \end{pmatrix}=
        \begin{pmatrix}
            \frac{1}{2} & 0 & \frac{1}{2} & 0 & \frac{1}{2} & 0 & \frac{1}{2} & 0\\
            0 & \frac{1}{2} & 0 & \frac{1}{2} & 0 & \frac{1}{2} & 0 & \frac{1}{2}\\
            \frac{1}{2\sqrt{3}} & \frac{1}{\sqrt{6}}e^{i\pi/4} & -\frac{1}{2\sqrt{3}} & -\frac{1}{\sqrt{6}}e^{i3\pi/4} & -\frac{1}{2\sqrt{3}} &  \frac{1}{\sqrt{6}}e^{i3\pi/4} & \frac{1}{2\sqrt{3}} &  -\frac{1}{\sqrt{6}}e^{i\pi/4}\\
            \frac{1}{\sqrt{6}}e^{-i\pi/4}  &-\frac{1}{2\sqrt{3}} & -\frac{1}{\sqrt{6}}e^{-i3\pi/4} & \frac{1}{2\sqrt{3}} & \frac{1}{\sqrt{6}}e^{-i3\pi/4}  & \frac{1}{2\sqrt{3}} & -\frac{1}{\sqrt{6}}e^{-i\pi/4} & -\frac{1}{2\sqrt{3}}\\
            \frac{1}{2\sqrt{3}}i & -\frac{3+\sqrt{3}}{6\sqrt{2}}e^{i3\pi/4} & -\frac{1}{2\sqrt{3}}i & \frac{3-\sqrt{3}}{6\sqrt{2}}e^{i\pi/4} & -\frac{1}{2\sqrt{3}}i & -\frac{3-\sqrt{3}}{6\sqrt{2}}e^{i\pi/4} & \frac{1}{2\sqrt{3}}i & \frac{3+\sqrt{3}}{6\sqrt{2}}e^{i3\pi/4}\\
            \frac{3-\sqrt{3}}{6\sqrt{2}}e^{i\pi/4} &  -\frac{1}{2\sqrt{3}}i & -\frac{3+\sqrt{3}}{6\sqrt{2}}e^{i3\pi/4} & \frac{1}{2\sqrt{3}}i & \frac{3+\sqrt{3}}{6\sqrt{2}}e^{i3\pi/4} & \frac{1}{2\sqrt{3}}i & -\frac{3-\sqrt{3}}{6\sqrt{2}}e^{i\pi/4} & -\frac{1}{2\sqrt{3}}i\\
            \frac{3+\sqrt{3}}{6\sqrt{2}} & \frac{1}{2\sqrt{3}}e^{i\pi/4} & -\frac{3-\sqrt{3}}{6\sqrt{2}}i & -\frac{1}{2\sqrt{3}}e^{i\pi/4} & \frac{3-\sqrt{3}}{6\sqrt{2}}i &-\frac{1}{2\sqrt{3}}e^{i\pi/4} & -\frac{3+\sqrt{3}}{6\sqrt{2}} & \frac{1}{2\sqrt{3}}e^{i\pi/4}\\
            \frac{1}{2\sqrt{3}}e^{i3\pi/4} & -\frac{3-\sqrt{3}}{6\sqrt{2}} & -\frac{1}{2\sqrt{3}}e^{i3\pi/4} & -\frac{3+\sqrt{3}}{6\sqrt{2}}i & -\frac{1}{2\sqrt{3}}e^{i3\pi/4} & \frac{3+\sqrt{3}}{6\sqrt{2}}i & \frac{1}{2\sqrt{3}}e^{i3\pi/4} & \frac{3-\sqrt{3}}{6\sqrt{2}}
        \end{pmatrix}
        \begin{pmatrix}
            |\psi_{1,\uparrow}\rangle\\
            |\psi_{1,\downarrow}\rangle\\
            |\psi_{2,\uparrow}\rangle\\
            |\psi_{2,\downarrow}\rangle\\
            |\psi_{3,\uparrow}\rangle\\
            |\psi_{3,\downarrow}\rangle\\
            |\psi_{4,\uparrow}\rangle\\
            |\psi_{4,\downarrow}\rangle\\
        \end{pmatrix}$.
        }
    \end{align}
    \end{widetext}
Projecting onto the $\overline{\Gamma}_{10}$ subspace, we obtain the degenerate perturbation matrix of the hopping Hamiltonian near the $\Gamma$ point,
\begin{align}
        \mathcal{H}^{\overline{\Gamma}_{10}}_{\mathrm{hopping},\alpha\beta}=
            \langle \psi^{\overline{\Gamma}_{10}}_\alpha|\mathcal{H}_{\mathrm{hopping}}|\psi^{\overline{\Gamma}_{10}}_\beta\rangle,
    \end{align}
    where $\alpha,\beta=1,2,3,4$. Explicitly, we obtain
    \begin{widetext}
        \begin{align}
            \mathcal{H}^{\overline{\Gamma}_{10}}_{\mathrm{hopping}}(\mathbf{k})=&\frac{2t'-t}{3}(3+\cos k_x\cos k_y+\cos k_y\cos k_z+\cos k_z\cos k_x)\mathbb{1}\nonumber\\
            -&\frac{t+t'}{3}(\cos k_x-\cos k_y)\cos k_z\left(J_x^2-J_y^2\right)+\frac{t+t'}{9}[2\cos k_x\cos k_y-\cos k_z(\cos k_x+\cos k_y)]\left(2J_z^2-J_x^2-J_y^2\right)\nonumber\\
            +&\sum_{\alpha\beta\gamma}\left(\frac{t'-20t}{9}J_\alpha+\frac{8t-2t'}{9}J_\alpha^3\right)\sin k_\alpha(\cos k_\beta-\cos k_\gamma)\nonumber\\
            +&\sum_{\alpha\beta\gamma}\frac{2t'-t}{3}\sin k_\beta\sin k_\gamma (J_\beta J_\gamma+J_\gamma J_\beta),
        \end{align}
    \end{widetext}
    where $\sum_{\alpha\beta\gamma}$ denotes the summation over three permutations $(\alpha,\beta,\gamma)=(x,y,z)$, $(\alpha,\beta,\gamma)=(y,z,x)$, and $(\alpha,\beta,\gamma)=(z,x,y)$.
    Expanding $\mathcal{H}^{\overline{\Gamma}_{10}}_{\mathrm{hopping}}(\mathbf{k})$ to its quadratic level and dropping the constant energy shift, we obtain
    \begin{widetext}
        \begin{align}
            \mathcal{H}^{\overline{\Gamma}_{10},(2)}_{\mathrm{hopping}}(\mathbf{k})=&\frac{t-2t'}{3}(k_x^2+k_y^2+k_z^2-6)\mathbb{1}\nonumber\\
            +&\frac{t+t'}{18}\left[3(k_x^2-k_y^2)(J_x^2-J_y^2)+(2k_z^2-k_x^2-k_y^2)(2J_z^2-J_x^2-J_y^2)\right]\nonumber\\
            -&\frac{t-2t'}{3}\left[k_yk_z(J_yJ_z+J_zJ_y)+ k_z k_x (J_xJ_z+J_zJ_x)+k_x k_y(J_xJ_y+J_yJ_x)\right]
        \end{align}
    \end{widetext}
    in agreement with the $A_{1g}$, $E_{g}$, and $T_{2g}$ coupling symmetry channels identified above.

    \subsection{Interaction term}
We start from the onsite Hubbard interaction in the microscopic Hamiltonian [Eq.~\eqref{eqn:tightbinding}],
\begin{align}
    \mathcal{H}_{\mathrm{int}}
    =
    U \sum_i n_{i\uparrow} n_{i\downarrow}.
\end{align}
The mean-field decoupling produces a local Zeeman-like term on each sublattice,
\begin{align}
    \mathcal{H}_{\mathrm{MF}}
    =
    -2US \sum_{\mathbf r,\alpha}
    \mathbf s_{\mathbf r,\alpha} \cdot \hat{\mathbf e}_\alpha,
\end{align}
where $\mathbf s_{\mathbf r,\alpha}$ denotes the local $j_{\mathrm{eff}}=1/2$ spin operator at site $\alpha$ in unit cell $\mathbf r$, and $\hat{\mathbf e}_\alpha$ are the AIAO unit vectors defined in Eq.~\eqref{eqn:AIAO}.

Using the basis transformation \eqref{eqn:projectU}, which maps the local spins on the four sublattices to the symmetry-adapted basis, the mean-field interaction can be written in matrix form as
 \begin{align}
        \mathcal{H}_{\mathrm{int}}^{\mathrm{MF}}=US
        \begin{pmatrix}
             0 & 0 &-1 & 0 & 0 & 0 & 0 & 0\\
             0 & 0 & 0 &-1 & 0 & 0 & 0 & 0\\
            -1 & 0 & 0 & 0 & 0 & 0 & 0 & 0\\
             0 &-1 & 0 & 0 & 0 & 0 & 0 & 0\\
             0 & 0 & 0 & 0 & 1 & 0 & 0 & 0\\
             0 & 0 & 0 & 0 & 0 & 1 & 0 & 0\\
             0 & 0 & 0 & 0 & 0 & 0 &-1 & 0\\
             0 & 0 & 0 & 0 & 0 & 0 & 0 &-1\\
        \end{pmatrix}.
    \end{align}
The upper and lower $4\times 4$ blocks correspond to the $\overline{\Gamma}_6 \oplus \overline{\Gamma}_7$ and $\overline{\Gamma}_{10}$ sectors, respectively. 

Focusing on the $\overline{\Gamma}_{10}$ subspace, the lower $4\times 4$ block can be expressed in terms of spin-$3/2$ operators [Eq.(\ref{Jmatrices})]. It is proportional to the cubic combination
\begin{align}
    J_xJ_yJ_z + J_zJ_yJ_x.\nonumber
\end{align}
Accordingly, the projected interaction Hamiltonian takes the form
\begin{align}
    \mathcal{H}^{\overline{\Gamma}_{10}}_{\mathrm{MF}}
    =
    \frac{2\sqrt{3}}{3}US(J_xJ_yJ_z + J_zJ_yJ_x),
\end{align}
where the prefactor $2\sqrt{3}/3$ is determined by matching the matrix elements of the lower $4\times4$ block to the corresponding spin-$3/2$ operator expression.

\section{Derivation of the coupling Hamiltonian coupled with the the $A_{2g}^-$ order parameter}
\label{app:C}

In this Appendix, we provide the technical details underlying the
symmetry-based construction of the effective low-energy
Hamiltonian in the presence of a condensed $A_{2g}^{-}$ order
parameter. Our goal is to derive the most general form of the coupling
within the low-energy $\overline{\Gamma}_{10}$ (denoted as $\Gamma_8^{+}$ or $\overline{F}_g$ in other conventions) manifold that is consistent with
the symmetry constraints of the magnetic point group. This construction
makes explicit how the $A_{2g}^{-}$ order lifts the fourfold degeneracy of
the  $\overline{\Gamma}_{10}$  states via the projected Hamiltonian introduced in the main text
[Eqs.~\eqref{Hint} and \eqref{HintA2g}].

We identify all electronic operators acting within the $\overline{\Gamma}_{10}$
subspace that transform according to the $A_{2g}^{-}$ irreducible
representation of the magnetic point group and are therefore allowed to
couple linearly to the AIAO order parameter. Since the $\overline{\Gamma}_{10}$
manifold is four dimensional, the effective coupling is represented
by a $4\times4$ Hermitian matrix acting on the $\overline{\Gamma}_{10}$ basis
states.

We begin from the most general $4\times4$ Hermitian matrix,
    \begin{align}
        \mathcal{H}_{\mathrm{coupl}}^{A_{2g}^-}=
        \begin{pmatrix}
            h_1         & h_5+ih_6       & h_7+ih_8       & h_9+ih_{10}\\
            h_5-ih_6    & h_2            & h_{11}+ih_{12} & h_{13}+ih_{14}\\
            h_7-ih_8    & h_{11}-ih_{12} & h_3            & h_{15}+ih_{16}\\
            h_9-ih_{10} & h_{13}-ih_{14} & h_{15}-ih_{16} & h_4
        \end{pmatrix},
    \end{align}
which contains sixteen real, $\mathbf{k}$-dependent parameters
$h_1,\dots,h_{16}$. Each coefficient may be systematically expanded in
powers of momentum as
     $h_{i}=h_i^{(0)}+h_i^{(1)}(\mathbf{k})+h_i^{(2)}(\mathbf{k})+\cdots$,
corresponding to the decomposition $\mathcal{H}_{\mathrm{coupl}}^{A_{2g}^-}=\mathcal{H}_{\mathrm{coupl}}^{A_{2g}^-,(0)}+\mathcal{H}_{\mathrm{coupl}}^{A_{2g}^-,(1)}+\mathcal{H}_{\mathrm{coupl}}^{A_{2g}^-,(2)}+\cdots$. 
At each order in momentum, the number of independent parameters is
constrained by enforcing the transformation properties of the $A_{2g}^{-}$ irrep under the symmetry operations of the magnetic point
group. In the remainder of this Appendix, we explicitly carry out this
symmetry reduction and derive the allowed form of
 $\mathcal{H}_{\mathrm{coupl}}^{A_{2g}^-}$ up to quadratic order in
$\mathbf{k}$.
    
    \subsection{The zeroth order term}
    The action of unitary generators on $\mathcal{H}_{\mathrm{coupl}}^{A_{2g}^-}$ consists of the $\overline{\Gamma}_{10}$ matrix representations of the magnetic point group $m\bar{3}m1'$ which can be found from the Bilbao Crystallographic Server \cite{Elcoro2017}. For the zeroth order $\mathcal{H}_{\mathrm{coupl}}^{A_{2g}^-,(0)}$, we only need matrices for three generators: $C_{2z}$ ($C_{2}$ rotation matrix along the $z$ axis),  $C_{2,[110]}$ ($C_{2}$ rotation matrix along the $[110]$ axis), and $C_{3,[111]}$ ($C_{3}$ rotation matrix along the $[111]$ axis). These matrices are 
    \begin{widetext}
        \begin{align}
            C_{2z}=
            \begin{pmatrix}
                -i & 0 & 0 & 0\\
                0  & i & 0 & 0\\
                0  & 0 & i & 0\\
                0  & 0 & 0 & -i
            \end{pmatrix},\quad
            C_{2,[110]}=
            \begin{pmatrix}
                0 & 0 & -1 & 0\\
                0 & 0 &  0 &-1\\
                1 & 0 &  0 & 0\\
                0 & 1 &  0 & 0
            \end{pmatrix},\quad
            C_{3,[111]}=
            \begin{pmatrix}
                \frac{\sqrt{2}}{2}e^{\frac{5\pi}{12}i} & \frac{\sqrt{2}}{2}e^{-\frac{\pi}{12}i} & 0 & 0\\
                \frac{\sqrt{2}}{2}e^{\frac{5\pi}{12}i} & \frac{\sqrt{2}}{2}e^{\frac{11\pi}{12}i} & 0 & 0\\
                0  & 0 & \frac{\sqrt{2}}{2}e^{-\frac{5\pi}{12}i} & \frac{\sqrt{2}}{2}e^{\frac{7\pi}{12}i}\\
                0  & 0 & \frac{\sqrt{2}}{2}e^{-\frac{11\pi}{12}i} & \frac{\sqrt{2}}{2}e^{-\frac{11\pi}{12}i}
            \end{pmatrix}.
        \end{align}
    \end{widetext}
    Moreover, the zeroth order $\mathcal{H}_{\mathrm{coupl}}^{A_{2g}^-,(0)}$ is independent of $\mathbf{k}$, so symmetry operations represented by the above matrices are directly applied to $\mathcal{H}_{\mathrm{coupl}}^{A_{2g}^-,(0)}$ without additional operations in the momentum space. Therefore, we can reduce the $16$ free $\mathbf{k}$-independent free variables ($h_1^{(0)}\dots,h_{16}^{(0)}$) by projecting onto the $A_{2g}^-$ irrep:
    \begin{align}
        C_{2z}\mathcal{H}_{\mathrm{coupl}}^{A_{2g}^-,(0)}C_{2z}^{-1}&=\mathcal{H}_{\mathrm{coupl}}^{A_{2g}^-,(0)},\nonumber\\
        C_{2,[110]}\mathcal{H}_{\mathrm{coupl}}^{A_{2g}^-,(0)}C_{2,[110]}^{-1}&=-\mathcal{H}_{\mathrm{coupl}}^{A_{2g}^-,(0)},\label{eqn:h0}\\
        C_{3,[111]}\mathcal{H}_{\mathrm{coupl}}^{A_{2g}^-,(0)}C_{3,[111]}^{-1}&=\mathcal{H}_{\mathrm{coupl}}^{A_{2g}^-,(0)}.\nonumber
    \end{align}
    The first equality reduces $h_5^{(0)}=h_6^{(0)}=h_7^{(0)}=h_8^{(0)}=h_{13}^{(0)}=h_{14}^{(0)}=h_{15}^{(0)}=h_{16}^{(0)}=0$. Then we can plug the reduced $\mathcal{H}_{\mathrm{int},A_{2g}^-}^{(0)}$ into the second equality. This gives us $h_3=-h_1$, $h_4=-h_2$, $h_{12}=-h_{10}$, and $h_{11}=h_9$. Plugging them into the last equality in (\ref{eqn:h0}) produces $h_2=h_1$ and $h_9=h_{10}=0$. Hence we obtain the zeroth order term
    \begin{align}
        {H}_{\mathrm{coupl}}^{A_{2g}^-,(0)}=
        \begin{pmatrix}
                h_1^{(0)} & 0         & 0         & 0\\
                0         & h_1^{(0)} & 0         & 0\\
                0         & 0         &-h_1^{(0)} & 0\\
                0         & 0         & 0         &-h_1^{(0)}
        \end{pmatrix}.\label{eqn:h0expression}
    \end{align}
    We can explicitly check the above form of ${H}_{\mathrm{coupl}}^{A_{2g}^-,(0)}$ indeed transforms according to the $\mathrm{A_{2g}^-}$ irreducible representation after applying the operations from the remaining generators.

    \subsection{The linear and quadratic order terms}
    
    For any higher order $\mathbf{k}$-dependent ${H}_{\mathrm{coupl},{A_{2g}^-}}^{(m)}$ ($m\ge1$), each symmetry operation needs to transform $\mathbf{k}$ accordingly in the momentum space in addition to symmetry operation as matrix multiplications. Consider the inversion operation in $\overline{\Gamma}_{10}$
    \begin{align}
        \mathcal{P}=\begin{pmatrix}
            1 & 0 & 0 & 0\\
            0 & 1 & 0 & 0\\
            0 & 0 & 1 & 0\\
            0 & 0 & 0 & 1
        \end{pmatrix},
    \end{align}
    which acts as the $4\times 4$ identity matrix. Under inversion, we must transform $\mathbf{k}\rightarrow-\mathbf{k}$ after performing $\mathcal{P}\mathcal{H}_{\mathrm{coupl}}^{A_{2g}^-}\mathcal{P}^{-1}$. Since the ${A_{2g}^-}$ irrep requires $\mathcal{H}_{\mathrm{coupl}}^{A_{2g}^-}$ to be invariant under inversion, all the odd orders of $\mathcal{H}_{\mathrm{coupl}}^{A_{2g}^-}$ must vanish, so we obtain $\mathcal{H}_{\mathrm{coupl}}^{A_{2g}^-,(1)}=0$.

    For the quadratic order $\mathcal{H}_{\mathrm{coupl}}^{A_{2g}^-,(2)}$, we expand $h_i^{(2)}(\mathbf{k})$ in all quadratic combinations of $\mathbf{k}$:
    \begin{align}
        h_i^{(2)}(\mathbf{k})=a_i k_x^2+b_i k_y^2+c_ik_z^2+d_i k_xk_y+e_i k_yk_z+f_i k_z k_x,
    \end{align}
    where $a_i$, $b_i$, $c_i$, $d_i$, $e_i$, and $f_i$ are $\mathbf{k}$-independent free variables. The reduction procedure is similar to \eqref{eqn:h0}, but combined with the corresponding transformations performed on $\mathbf{k}$. We start with
    \begin{align}
        C_{2z}\mathcal{H}_{\mathrm{coupl}}^{A_{2g}^-,(2)}(-k_x,-k_y,k_z)C_{2z}^{-1}&=\mathcal{H}_{\mathrm{coupl}}^{A_{2g}^-,(2)}(\mathbf{k}),
    \end{align} 
    where the $\mathbf{k}$-dependence in $\mathcal{H}_{\mathrm{coupl}}^{A_{2g}^-,(2)}(-k_x,-k_y,k_z)$  denotes performing the transformation $(k_x,k_y,k_z)\rightarrow(-k_x,-k_y,k_z)$ after the matrix multiplication. This gives $a_{5,6,7,8,13,14,15,16}=0$, $b_{5,6,7,8,13,14,15,16}=0$, $c_{5,6,7,8,13,14,15,16}=0$, $d_{5,6,7,8,13,14,15,16}=0$, $e_{1,2,3,4,9,10,11,12}=0$, $f_{1,2,3,4,9,10,11,12}=0$. Next, we use
    \begin{align}
        C_{2,[110]}\mathcal{H}_{\mathrm{coupl}}^{A_{2g}^-,(2)}(k_y,k_x,-k_z)C_{2,[110]}^{-1}&=-\mathcal{H}_{\mathrm{coupl}}^{A_{2g}^-,(2)}(\mathbf{k}).
    \end{align} 
    This reduces $a_1=-b_3$, $a_2=-b_4$, $a_{3}=-b_{1}$, $a_{4}=-b_{2}$, $a_{9}=b_{11}$, $a_{10}=-b_{12}$, $a_{11}=b_{9}$, $a_{12}=-b_{10}$, $c_1=-c_3$, $c_2=-c_4$, $c_9=c_{11}$, $c_{10}=-c_{12}$, $d_1=-d_3$, $d_2=-d_4$, $d_9=d_{11}$, $d_{10}=-d_{12}$, $e_{5}=f_{15}$, $e_{6}=f_{16}$, $e_{7}=-f_{7}$, $e_{8}=f_{8}$, $e_{13}=-f_{13}$, $e_{14}=f_{14}$. Then we apply
    \begin{align}
        C_{3,[111]}\mathcal{H}_{\mathrm{coupl}}^{A_{2g}^-,(2)}(k_y,k_z,k_x)C_{3,[111]}^{-1}&=\mathcal{H}_{\mathrm{coupl}}^{A_{2g}^-,(2)}(\mathbf{k}).
    \end{align}
    This produces $b_1=-c_4$, $b_2=-c_4$, $b_3=c_4$, $b_4=c_4$, $b_9=c_{12}(\sqrt{3}-1)/2$, $b_{10}=c_{12}(1+\sqrt{3})/2$, $b_{11}=-c_{12}(1+\sqrt{3})/2$, $b_{12}=c_{12}(\sqrt{3}-1)/2$, $c_3=c_4$, $c_{11}=c_{12}$, $d_3=-f_{15}$, $d_4=f_{15}$, $d_{11}=(1-\sqrt{3})f_{14}$, $d_{12}=(\sqrt{3}-1)f_{14}$, $f_{5,16}=0$, $f_6=-f_{15}$, $f_7=f_{14}$, $f_{8}=(\sqrt{3}-2)f_{14}$, $f_{13}=(2-\sqrt{3})f_{14}$. Finally, we consider the application of the time reversal operator 
    \begin{align}
        \mathcal{K}\mathcal{H}_{\mathrm{coupl}}^{A_{2g}^-,(2)}(-k_x,-k_y,-k_z)\mathcal{K}^{-1}&=-\mathcal{H}_{\mathrm{coupl}}^{A_{2g}^-,(2)}(\mathbf{k}),
    \end{align}
    with the unitary part of $\mathcal{K}$ in $\overline{\Gamma}_{10}$
    \begin{align}
        \begin{pmatrix}
            0 & 0 & e^{-i\pi/4} & 0\\
            0 & 0 & 0 & e^{i\pi/4}\\
            e^{i\pi/4} & 0 & 0 & 0\\
            0 & e^{-i\pi/4} & 0 & 0
        \end{pmatrix}.
    \end{align}
    This further reduces $c_{12}=0$. Therefore, after rescaling the remaining constants, we obtain 
    \begin{widetext}
    \begin{align}
    &\mathcal{H}_{\mathrm{coupl}}^{A_{2g}^-,(2)}(\mathbf{k})=\nonumber\\
        &\scalebox{0.9}{$\begin{pmatrix}
            -c_4 k^2 +f_{15} k_x k_y & f_{15}(-i k_x k_z+k_yk_z) & f_{14}(e^{-i \pi/12}k_z k_x+e^{-i 11\pi/12}k_y k_z) & f_{14}e^{-i 3\pi/4}k_x k_y\\
            f_{15}(i k_x k_z+k_yk_z) & -c_4 k^2 -f_{15} k_x k_y  & f_{14}e^{-i3\pi/4}k_x k_y & f_{14}(e^{i 5\pi/12}k_z k_x+e^{i 7\pi/12}k_y k_z)\\
            f_{14}(e^{i \pi/12}k_z k_x+e^{i 11\pi/12}k_y k_z) & f_{14}e^{-i 3\pi/4}k_x k_y & c_4 k^2 - f_{15} k_x k_y & f_{15} (-ik_yk_z+k_xk_z)\\
            f_{14}e^{i 3\pi/4}k_x k_y & f_{14}(e^{-i 5\pi/12}k_z k_x+e^{-i 7\pi/12}k_y k_z) & f_{15} (ik_yk_z+k_xk_z) & c_4 k^2 + f_{15} k_x k_y
        \end{pmatrix}$},\label{eqn:h2expression}
    \end{align}
    \end{widetext}
    where $k^2\equiv k_x^2+k_y^2+k_z^2$, with three free real variables $c_4$, $f_{14}$, and $f_{15}$. 
    \subsection{$J=3/2$ operator representation}

    To verify that the symmetry-allowed Hamiltonian $\mathcal{H}^{A_{2g}^-,(0)}_{\mathrm{coupl}}+\mathcal{H}^{A_{2g}^-,(2)}_{\mathrm{coupl}}(\mathbf{k})$ transforms under the $A_{2g}^-$ irrep, we rewrite it in terms of the spin-$3/2$ operators $J_x$, $J_y$, and $J_z$ in the basis of \eqref{eqn:3halfbasis}.
    
    From the product table for the irreps of $m\bar{3}m$, the $A_{2g}$ irrep arises from $T_{1g}\otimes T_{2g}$ and $A_{1g}\otimes A_{2g}$, both products of inversion-even irreps. The relevant $\mathbf{k}$-dependent basis functions are
    \begin{align}
        A_{1g}: \quad&  k_x^2+k_y^2+k_z^2,\nonumber\\
        T_{2g}: \quad&  (k_y k_z, k_x k_z,k_x k_y),\label{eqn:J2channels}
    \end{align} 
    and the spin-$3/2$ operator bases are
    \begin{align}
        A_{2g}: \quad& J_x J_y J_z+J_z J_y J_x,\nonumber\\
        T_{1g}: \quad& (J_x,J_y,J_z), \\
                \quad& (J_x^3,J_y^3,J_z^3).\nonumber
    \end{align}
    The $A_{2g}$ channel is then constructed by analogy with the spin-$1/2$ case: $k_y k_z \sigma_x+k_x k_z \sigma_y+k_x k_y \sigma_z$ in terms of the Pauli matrices $\sigma_x,\sigma_y,\sigma_z$ \cite{Fernandes2024}. Then we obtain $k_y k_z J_x+k_x k_z J_y+k_x k_y J_z$. The octupolar nature of spin-$3/2$ operators further contributes $k_y k_z J_x^3+k_x k_z J_y^3+k_x k_y J_z^3$, $J_x J_y J_z+J_z J_y J_x$ and $|\mathbf{k}|^2(J_x J_y J_z+J_z J_y J_x)$. All four terms are odd under time reversal, confirming the $A_{2g}^-$ label. Summing these channels with independent coefficients yields the symmetry-allowed low-energy Hamiltonian in Eq. \eqref{eqn:symmetryA2g}. Expanding explicitly, the constants in \eqref{eqn:h0expression} and \eqref{eqn:h2expression} are identified as
    \begin{align}
        h_1^{(0)}    &= \frac{\sqrt{3}}{2} \lambda_3,\nonumber\\
        c_{4}  &= \frac{\sqrt{3}}{2}\lambda_4,\nonumber\\
        f_{15} &=-\frac{4\lambda_1+13\lambda_2}{8},\\
        f_{14} &= \frac{4\lambda_1+7\lambda_2}{4}.\nonumber
    \end{align}

\section{Local magnetization density}\label{app:D}
       
        The local magnetization density can be computed from the local charge current and local spin density 
    \begin{align}
        \mathbf{m}(\mathbf{r})=\frac{1}{2}\mathbf{r}\times \mathbf{j}(\mathbf{r})+2 \psi^\dagger(\mathbf{r}) \mathbf{S}\psi(\mathbf{r}),
    \end{align}
    where the local charge current is defined by   \begin{align}
        \mathbf{j}(\mathbf{r})=-i\left(\psi^\dagger(\mathbf{r})\nabla\psi(\mathbf{r})-\nabla\psi^\dagger(\mathbf{r})\psi(\mathbf{r})\right).
    \end{align}
    Here, $\mathbf{S}$ denotes the vector of spin $1/2$ matrices, $(\sigma_x/2,\sigma_y/2,\sigma_z/2)$, and $\psi(\mathbf{r})$ is the spinor form of the single-electron wave function. $\mathbf{m}(\mathbf{r})$ is in units of the Bohr magneton $\mu_B$. For simplicity, we omit repeating the $\mathbf{r}$ dependence in the remaining expressions unless otherwise specified.
    
    In spherical coordinates, the local charge current density can be written as 
    \begin{align}
        \mathbf{j}=\mathrm{Im}\left(2\psi^\dagger\frac{\partial \psi}{\partial r}\hat{\mathbf{r}}+\frac{2}{r}\psi^\dagger\frac{\partial \psi}{\partial \theta}\hat{\boldsymbol{\theta}}+\frac{2}{r\sin\theta}\psi^\dagger\frac{\partial \psi}{\partial \phi}\hat{\boldsymbol{\phi}}\right),
    \end{align}
which makes explicit the separate contributions from the radial and angular variations of the electronic wave function. Since the orbital magnetization density is given by
$
\frac{1}{2}\,\mathbf r \times \mathbf j,
$
the radial component of the current does not contribute.
As a result, only the angular derivatives generate an orbital magnetic moment, yielding
\begin{align}
        \frac{1}{2}\mathbf{r}\times\mathbf{j}= \mathrm{Im}\left(\psi^\dagger\frac{\partial\psi}{\partial\theta}\hat{\boldsymbol{\phi}}-\frac{1}{\sin\theta}\psi^\dagger\frac{\partial\psi}{\partial \phi}\hat{\boldsymbol{\theta}}\right).
    \end{align}
Rewriting  the orbital contribution in Cartesian coordinates and adding the local spin density, we obtain the magnetization density
    \begin{widetext}
        \begin{align}
            m_x&=-\mathrm{Im}\left(\psi^\dagger\frac{\partial\psi}{\partial\theta}\right)\sin\phi-\mathrm{Im}\left(\psi^\dagger\frac{\partial\psi}{\partial\phi}\right)\cos\phi\cot\theta+\psi^\dagger\sigma_x\psi,\nonumber \\
            m_y&= \mathrm{Im}\left(\psi^\dagger\frac{\partial\psi}{\partial\theta}\right)\cos\phi-\mathrm{Im}\left(\psi^\dagger\frac{\partial\psi}{\partial\phi}\right)\sin\phi\cot\theta+\psi^\dagger\sigma_y\psi,\label{eqn:magnetization}\\
            m_z&= \mathrm{Im}\left(\psi^\dagger\frac{\partial\psi}{\partial\phi}\right)+\psi^\dagger\sigma_z\psi.\nonumber
        \end{align}
    \end{widetext}

Up to this point, the expressions for the local magnetization density $\mathbf m(\mathbf r)$ have been completely general. We now specialize these expressions to the AIAO mean-field state by adopting an explicit form for the electronic wave function. Specifically, we take $\psi(\mathbf r)$ to be a Bloch eigenstate $\psi_{\mathbf{k},\mu}(\mathbf r)$ of the mean-field Hamiltonian, expressed in the basis of localized $j_{\mathrm{eff}}=1/2$ orbitals on the four pyrochlore sublattices, such that
\begin{align}
        \psi_{\mathbf{k},\mu}(\mathbf{r})=\frac{1}{\sqrt{N}}\sum_{j,\nu}e^{i\mathbf{k}\cdot\mathbf{r}_j}(c_{\mathbf{k},\uparrow}^{\mu,\nu}\psi_{j,\nu,\uparrow}(\mathbf{r})+c_{\mathbf{k},\downarrow}^{\mu,\nu}\psi_{j,\nu,\downarrow}(\mathbf{r})),
    \end{align}
where $j$ labels unit cells, $\nu=1,\dots,4$ labels the sublattices, and $N$ is the number of unit cells. 
The coefficients $c_{\mathbf{k},\uparrow}^{\mu,\nu}$ and $c_{\mathbf{k},\downarrow}^{\mu,\nu}$ encode the AIAO mean-field texture and are obtained by diagonalizing the self-consistent mean-field Hamiltonian. They form the components of the corresponding eight-dimensional eigenvector,
 $(c_{\mathbf{k},\uparrow}^{\mu,1},c_{\mathbf{k},\downarrow}^{\mu,1},c_{\mathbf{k},\uparrow}^{\mu,2},c_{\mathbf{k},\downarrow}^{\mu,2},c_{\mathbf{k},\uparrow}^{\mu,3},c_{\mathbf{k},\downarrow}^{\mu,3},c_{\mathbf{k},\uparrow}^{\mu,4},c_{\mathbf{k},\downarrow}^{\mu,4})$, corresponding to the eigenvalue $\mathbf{\epsilon}_{\mathbf{k},\mu}$.

The localized $j_{\mathrm{eff}}=1/2$ orbitals at sublattice $\nu$ in unit cell $\mathbf r_j$ are constructed from the real cubic harmonics $d_{yz}$, $d_{zx}$, and $d_{xy}$ and can be written in spinor form as
    \begin{widetext}
    \begin{align}
        \psi_{j,\nu,\uparrow}&\propto\frac{1}{\sqrt{3}}\begin{pmatrix}
            0\\d_{yz}
        \end{pmatrix}+\frac{i}{\sqrt{3}}\begin{pmatrix}
            0\\d_{zx}
        \end{pmatrix}+\frac{1}{\sqrt{3}}\begin{pmatrix}
            d_{xy}\\0
        \end{pmatrix}=\frac{1}{4}\sqrt{\frac{5}{\pi}}\begin{pmatrix}
            \sin^2\theta_{j,\nu}\sin2\phi_{j,\nu}\\i\sin2\theta_{j,\nu} e^{-i\phi_{j,\nu}}
        \end{pmatrix},\\
        \psi_{j,\nu,\downarrow}&\propto\frac{1}{\sqrt{3}}\begin{pmatrix}
            d_{yz}\\0
        \end{pmatrix}-\frac{i}{\sqrt{3}}\begin{pmatrix}
            d_{zx}\\0
        \end{pmatrix}-\frac{1}{\sqrt{3}}\begin{pmatrix}
            0\\d_{xy}
        \end{pmatrix}
        =-\frac{1}{4}\sqrt{\frac{5}{\pi}}\begin{pmatrix}
            i\sin2\theta_{j,\nu} e^{i\phi_{j,\nu}}\\\sin^2\theta_{j,\nu}\sin2\phi_{j,\nu}
        \end{pmatrix}.
    \end{align}
\end{widetext}
     Here, the spherical coordinates $(\theta_{j,\nu},\phi_{j,\nu})$ are defined with respect to an origin centered at each sublattice site.
    Assuming the majority of the magnetization is confined in the vicinity of the sublattice site, we neglect cross terms involving different unit cells and sublattices when evaluating expressions such as $\psi_{\mathbf{k},\mu}^\dagger \partial_\theta \psi_{\mathbf{k},\mu}$. 
    With this local (on-site) approximation, we define the projected wave function
    $\psi_{\mathbf{k},\mu}^{j,\nu}=c_{\mathbf{k},\uparrow}^{\mu,\nu}\psi_{j,\nu,\uparrow}+c_{\mathbf{k},\downarrow}^{\mu,\nu}\psi_{j,\nu,\downarrow}$. From this point on, we fix a given sublattice site $(j,\nu)$ and suppress the corresponding subscripts on the angular variables, writing $(\theta,\phi)\equiv(\theta_{j,\nu},\phi_{j,\nu})$ for notational simplicity. With this convention, we compute the individual  orbital and spin contributions terms entering
 Eq.~\eqref{eqn:magnetization}: 
    \begin{widetext}
        \begin{align}
            \mathrm{Im}\left({\psi_{\mathbf{k},\mu}^{j,\nu}}^\dagger\frac{\partial\psi_{\mathbf{k},\mu}^{j,\nu}}{\partial\theta}\right)&=\frac{5}{4\pi}\sin^2\theta\sin2\phi\left(\mathrm{Re}(c_{\mathbf{k},\uparrow}^{\mu,\nu}{c_{\mathbf{k},\downarrow}^{\mu,\nu}}^{*})\cos\phi+\mathrm{Im}(c_{\mathbf{k},\uparrow}^{\mu,\nu}{c_{\mathbf{k},\downarrow}^{\mu,\nu}}^{*})\sin\phi\right),\nonumber\\
            \mathrm{Im}\left({\psi_{\mathbf{k},\mu}^{j,\nu}}^\dagger\frac{\partial\psi_{\mathbf{k},\mu}^{j,\nu}}{\partial\phi}\right)&=\frac{5}{4\pi}\sin^2\theta\cos\theta\left((|c_{\mathbf{k},\downarrow}^{\mu,\nu}|^2-|c_{\mathbf{k},\uparrow}^{\mu,\nu}|^2)\cos\theta+2\sin\theta \left(\mathrm{Re}(c_{\mathbf{k},\uparrow}^{\mu,\nu}{c_{\mathbf{k},\downarrow}^{\mu,\nu}}^{*})\cos^3\phi-\mathrm{Im}(c_{\mathbf{k},\uparrow}^{\mu,\nu}{c_{\mathbf{k},\downarrow}^{\mu,\nu}}^{*})\sin^3\phi\right)\right),
        \end{align}
        and
        \begin{align}
            {\psi_{\mathbf{k},\mu}^{j,\nu}}^\dagger\sigma_x\psi_{\mathbf{k},\mu}^{j,\nu}&=-\frac{5}{2\pi}\sin^2\theta\left(\mathrm{Re}(e^{-2i\phi}c_{\mathbf{k},\uparrow}^{\mu,\nu}{c_{\mathbf{k},\downarrow}^{\mu,\nu}}^{*})\cos^2\theta+((|c_{\mathbf{k},\downarrow}^{\mu,\nu}|^2-|c_{\mathbf{k},\uparrow}^{\mu,\nu}|^2)\cos\theta+\mathrm{Re}(c_{\mathbf{k},\uparrow}^{\mu,\nu}{c_{\mathbf{k},\downarrow}^{\mu,\nu}}^{*})\sin\theta\cos\phi)\sin\theta\cos\phi\sin^2\phi\right),\nonumber\\
            {\psi_{\mathbf{k},\mu}^{j,\nu}}^\dagger\sigma_y\psi_{\mathbf{k},\mu}^{j,\nu}&=-\frac{5}{2\pi}\sin^2\theta\left(\mathrm{Im}(e^{-2i\phi}c_{\mathbf{k},\uparrow}^{\mu,\nu}{c_{\mathbf{k},\downarrow}^{\mu,\nu}}^{*})\cos^2\theta-((|c_{\mathbf{k},\uparrow}^{\mu,\nu}|^2-|c_{\mathbf{k},\downarrow}^{\mu,\nu}|^2)\cos\theta+\mathrm{Im}(c_{\mathbf{k},\uparrow}^{\mu,\nu}{c_{\mathbf{k},\downarrow}^{\mu,\nu}}^{*})\sin\theta\sin\phi)\sin\theta\sin\phi\cos^2\phi\right),\nonumber\\
            {\psi_{\mathbf{k},\mu}^{j,\nu}}^\dagger\sigma_z\psi_{\mathbf{k},\mu}^{j,\nu}&=\frac{5}{64\pi}\sin^2\theta\left((|c_{\mathbf{k},\downarrow}^{\mu,\nu}|^2-|c_{\mathbf{k},\uparrow}^{\mu,\nu}|^2)(7+9\cos2\theta+2\sin^2\theta\cos4\phi)+16\mathrm{Re}(ie^{-i\phi}c_{\mathbf{k},\uparrow}^{\mu,\nu}{c_{\mathbf{k},\downarrow}^{\mu,\nu}}^{*})\sin2\theta\sin2\phi\right).
        \end{align}
    \end{widetext}
    Using the expression above, we can compute the local magnetization density at each sublattice site $\nu$ by summing the contributions from  all electrons in the system up to the Fermi surface
    \begin{widetext}
        \begin{align}
            m_{x,\nu}&=\frac{1}{N}\sum_{\mathbf{k}}^{k_F}\left(-\mathrm{Im}\left({\psi_{\mathbf{k},\mu}^{j,\nu}}^\dagger\frac{\partial\psi_{\mathbf{k},\mu}^{j,\nu}}{\partial\theta}\right)\sin\phi-\mathrm{Im}\left({\psi_{\mathbf{k},\mu}^{j,\nu}}^\dagger\frac{\partial\psi_{\mathbf{k},\mu}^{j,\nu}}{\partial\phi}\right)\cos\phi\cot\theta+{\psi_{\mathbf{k},\mu}^{j,\nu}}^\dagger\sigma_x\psi_{\mathbf{k},\mu}^{j,\nu}\right),\nonumber \\
            m_{y,\nu}&=\frac{1}{N}\sum_{\mathbf{k}}^{k_F}\left(\mathrm{Im}\left({\psi_{\mathbf{k},\mu}^{j,\nu}}^\dagger\frac{\partial\psi_{\mathbf{k},\mu}^{j,\nu}}{\partial\theta}\right)\cos\phi-\mathrm{Im}\left({\psi_{\mathbf{k},\mu}^{j,\nu}}^\dagger\frac{\partial\psi_{\mathbf{k},\mu}^{j,\nu}}{\partial\phi}\right)\sin\phi\cot\theta+{\psi_{\mathbf{k},\mu}^{j,\nu}}^\dagger\sigma_y\psi_{\mathbf{k},\mu}^{j,\nu}\right),\\
            m_{z,\nu}&=\frac{1}{N}\sum_{\mathbf{k}}^{k_F}\left(\mathrm{Im}\left({\psi_{\mathbf{k},\mu}^{j,\nu}}^\dagger\frac{\partial\psi_{\mathbf{k},\mu}^{j,\nu}}{\partial\phi}\right)+{\psi_{\mathbf{k},\mu}^{j,\nu}}^\dagger\sigma_z\psi_{\mathbf{k},\mu}^{j,\nu}\right).\nonumber
        \end{align}
    \end{widetext}

    To examine the symmetry properties of the local magnetization density, it is convenient to decompose each Cartesian component of the
     local magnetization density in terms of spherical harmonics. Accordingly, we expand
    \begin{align}\label{magnetizationcomponets}
        m_{\alpha,\mu}=\sum_{lm}C_{lm}^{\alpha,\mu}Y_{lm},
    \end{align}
    where the coefficients $C_{lm}^{\alpha,\mu}$
     quantify the contribution of the $(l,m)$ angular-momentum channel to the local magnetization profile. These coefficients are obtained by projecting onto the corresponding spherical harmonics,   
    \begin{align}
        C_{lm}^{\alpha,\mu}=\int d\Omega\  m_{\alpha,\mu} Y_{lm}^*,
    \end{align}
with $d\Omega=\sin\theta\,d\theta\,d\phi$. The resulting  distributions of the components of the local magnetization density (\ref{magnetizationcomponets})  are shown in Fig.~\ref{fig:magnetization-density} (a), with individual components for $l=0,2,4$ shown in 
Fig.~\ref{fig:magnetization-density} (b).
    
Beyond the angular decomposition, the multipolar character of the local magnetization on a given sublattice $\nu$ can be made explicit by defining the corresponding sublattice multipole moments, \cite{Hayami2018,Fernandes2024}
\begin{align}
    \mu_{lm}^{\nu}=\int d\mathbf{r}\, \mathbf{m}_{\nu}\cdot\nabla(r^l Y_{lm}^*),
\end{align}
 with $\mathbf{m}_{\nu}=(m_{x,\nu},m_{y,\nu},m_{z,\nu})$. This definition corresponds to the standard magnetic multipole expansion familiar from electromagnetism: the gradient ensures the correct transformation properties under rotations, while the integral projects the magnetization onto different angular-momentum channels.

\bibliography{references}
\end{document}